\documentclass{aa}  

\usepackage{graphicx}
\usepackage{txfonts}
\usepackage{natbib}
\usepackage{amssymb}
\usepackage{amsmath}
\usepackage{url}
\usepackage[section]{placeins}
\usepackage{multirow}
\usepackage{lscape}
\usepackage{wasysym}

\begin{document}
\title{Observations of nitrogen isotope fractionation \\ in deeply embedded protostars}

\author{S.~F.~Wampfler \inst{\ref{inst1},\ref{inst2}}
\and J.~K.~J\o{}rgensen \inst{\ref{inst2},\ref{inst1}} 
\and M.~Bizzarro  \inst{\ref{inst1}}
\and S.~E.~Bisschop \inst{\ref{inst1},\ref{inst2}}
}

\institute{
Centre for Star and Planet Formation, Natural History Museum of Denmark, University of Copenhagen, \O{}ster Voldgade 5-7, DK-1350 K\o{}benhavn K, Denmark\label{inst1}
\and
Niels Bohr Institute, University of Copenhagen, Juliane Maries Vej 30, DK-2100 K\o{}benhavn \O{}, Denmark\label{inst2}
}

\date{Received 7 March 2014 / Accepted 26 July 2014}

\titlerunning{Nitrogen isotope fractionation in embedded protostars}


\def\placefigureIRASspec{
\begin{figure*}
 \centering
 \resizebox{\hsize}{!}{\includegraphics{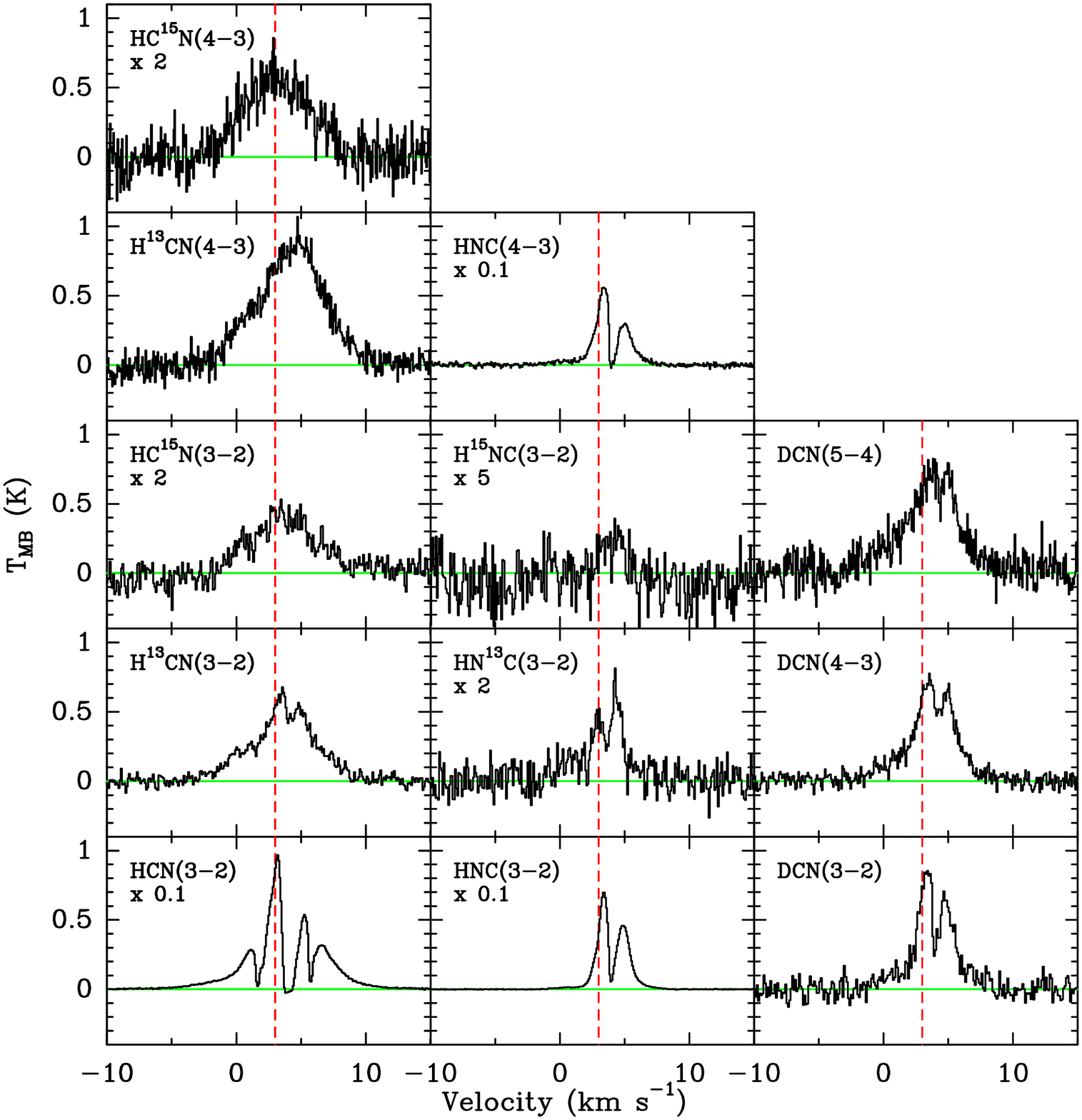}}
 \caption{Spectra of the HCN and HNC isotopologues observed with APEX toward IRAS~16293A. Where needed, the spectra have been scaled by the factors indicated in the individual panels.}
 \label{fig:iras16293_spectra}
\end{figure*}
}

\def\placefigureRCrAspec{
\begin{figure}[t]
 \centering
  \resizebox{\hsize}{!}{\includegraphics{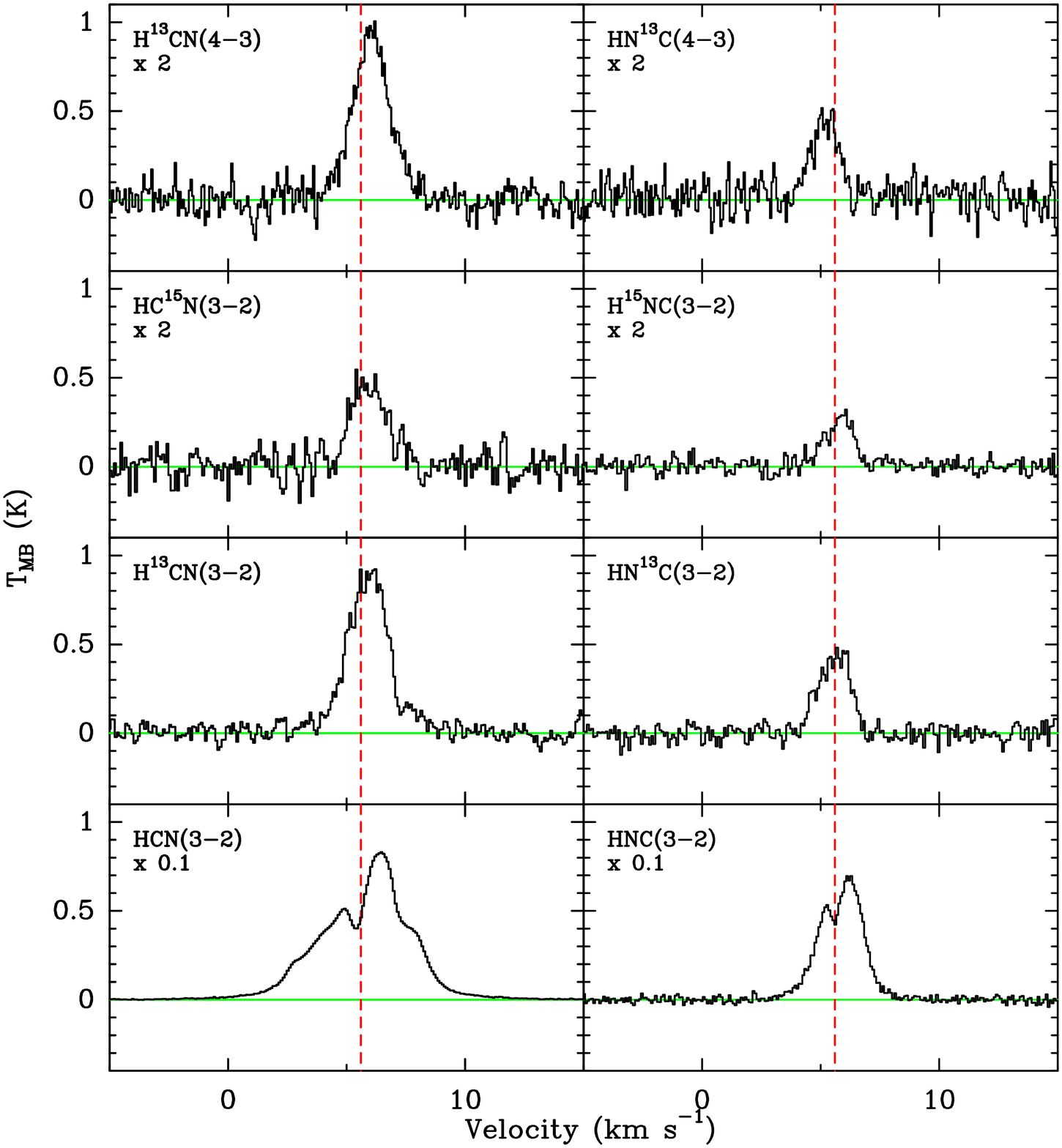}}
 \caption{Spectra of the HCN and HNC isotopologues observed with APEX toward R~CrA~IRS7B. Where needed, the spectra have been scaled by the factors indicated in the individual panels}
 \label{fig:rcrairs7b_spectra}
\end{figure}
}

\def\placefigureOMCspec{
\begin{figure}
 \centering
  \resizebox{\hsize}{!}{\includegraphics{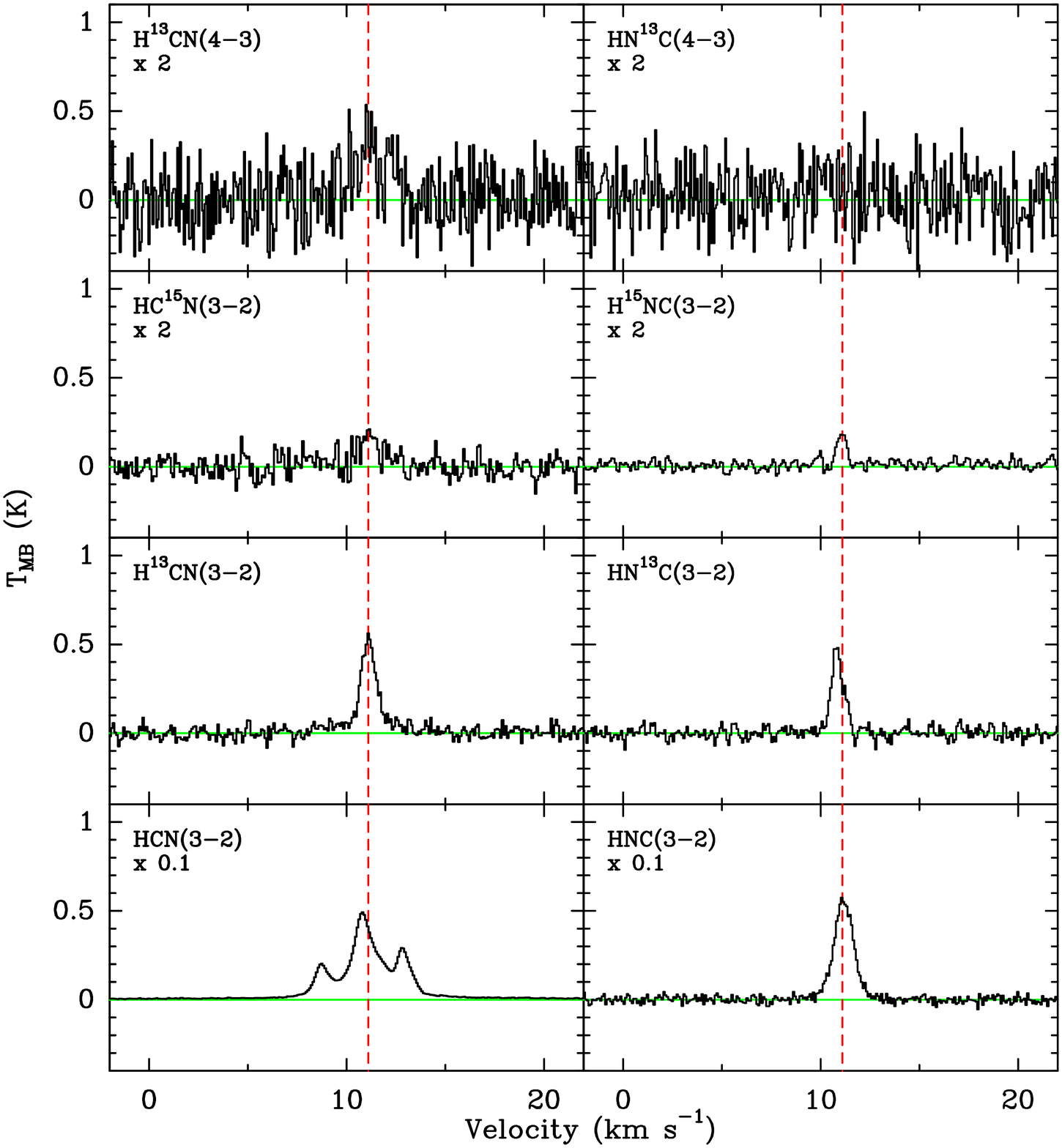}}
 \caption{Spectra of the HCN and HNC isotopologues observed with APEX toward OMC-3~MMS6. Where needed, the spectra have been scaled by the factors indicated in the individual panels}
 \label{fig:omc3mms6_spectra}
\end{figure}
}

\def\placefigureRatios{
\begin{figure*}
 \centering
 \resizebox{\hsize}{!}{\includegraphics[bb=0 0 430 175]{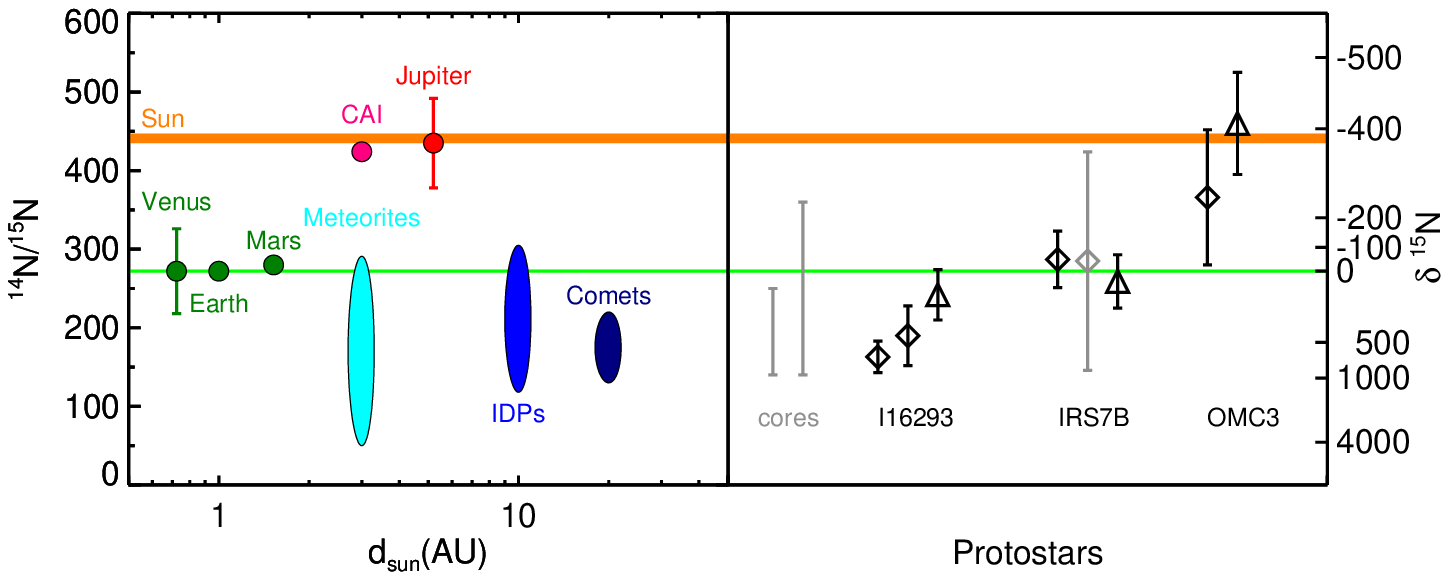}}
 \caption{Left panel: Illustration of the nitrogen isotope heterogeneity of the solar system, adapted from \citet{Meibom07,Mumma11}. Comets, meteorites, and interstellar dust particles (IDPs) are located over a range of distances from the Sun and the plotted $d_\mathrm{Sun}$ values should be regarded as an average value only. Right panel: ${}^{14}$N/${}^{15}$N ratios obtained from APEX measurements (black, this work) and ASTE \citep[grey,][]{Watanabe12}. Diamonds and triangles indicate values from HCN and HNC isotopologues, respectively. Error bars are $1\sigma$ for combined uncertainties on the calibration and the intensity. The gray bars (``cores'') indicate the prestellar core values from HCN by \citet{Hilyblant13a}.}
 \label{fig:nitrogen_ratios}
\end{figure*}
}

\def\placefigureSEST{
\begin{figure}
 \centering
 \resizebox{\hsize}{!}{\includegraphics[bb=31 122 576 632]{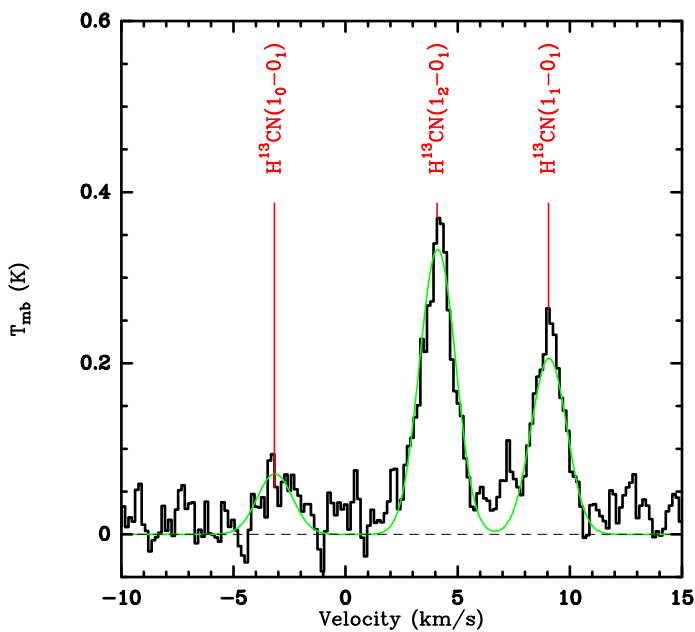}}
 \caption{H${}^{13}$CN(1-0) spectrum of IRAS~16293-2422 from the SEST telescope. The solid line shows the best fit from the CLASS software.}
 \label{fig:iras16293_sest}
\end{figure}
}

\def\placefigsourceprof{
\begin{figure}
 \centering
 \resizebox{\hsize}{!}{\includegraphics[angle=0,bb=0 0 249 294]{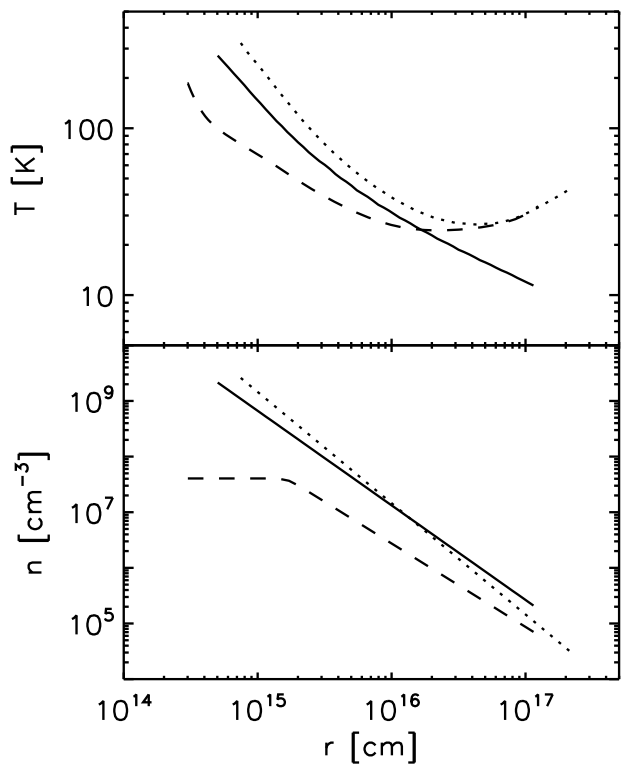}}
 \caption{Temperature and density as function of radius for IRAS~16293-2422 (solid line), OMC-3~MMS6 (dotted), and R~CrA~IRS7B (dashed).}
 \label{fig:sourceprofiles}
\end{figure}
}

\def\placefigtempratio{
\begin{figure}
 \centering
 \resizebox{\hsize}{!}{\includegraphics[angle=0,bb=0 0 249 175]{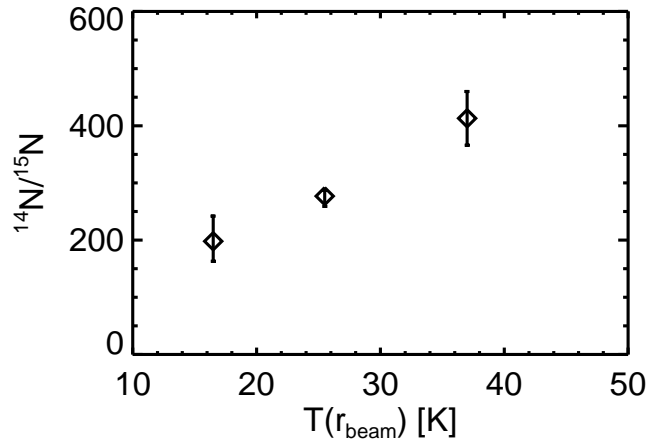}}
 \caption{Average ${}^{14}$N/${}^{15}$N ratios of the three young stellar objects plotted versus their outer envelope temperature at the projected beam radius. Error bars reflect the spread in measured ${}^{14}$N/${}^{15}$N ratios.}
 \label{fig:temp_ratio}
\end{figure}
}


\def\placetableprevnew{
\begin{table*}
\caption{Literature values of the ${}^{14}$N/${}^{15}$N ratio from low-mass prestellar cores and protostars (top). For comparison, the most nearby sources from the galactic survey by \citet{Adande12} are listed as well (bottom).} 
\centering
\begin{tabular}{lccccccl}
\hline \hline\noalign{\smallskip}
Source          & HCN                         & HNC           & CN               & NH$_3$                    & NH$_2$D         & N$_2$H$^{+}$ & Remarks\\
\hline\noalign{\smallskip}
L1544           & $140-360{}^{a}$             &               & $510\pm70{}^{b}$ &                           & $>700{}^{c}$    & $1000\pm200{}^{d}$ & prestellar core \\
                & $\phantom{0}69-154{}^{e}$   & $>27{}^{e}$   &                  &                           &                 &  &  \\
L1498           & $>813{}^{f}$                & $>90{}^{e}$   & $476\pm70{}^{b}$ &                           &                 &                    & prestellar core \\ 
L183            & $140-250{}^{a}$             &               &                  &                           &                 &                    & prestellar core \\ 
L1521E          & $151\pm\phantom{0}16{}^{f}$ &               &                  &                           &                 &                    & prestellar core \\
L1521F          & $>51{}^{e}$                 & $24-31{}^{e}$ &                  &                           &                 &                    & prestellar core \\        \\
Barnard~1       & $330^{+60}_{-50}{}^{g}$ & $225^{+75}_{-45}{}^{g}$ & $290^{+160}_{-80}{}^{g}$ & $300^{+55}_{-40}{}^{g}$ & $230^{+105}_{-55}{}^{g}$ &   $400^{+100}_{-65}{}^{g}, >600{}^{g}$                 &                  \\
NGC1333~IRAS4   &                             &               &                  &                           & $>270{}^{c}$    &                    & protostar        \\       
NGC1333~DCO$^+$ &                             &               &                  &                           & $360^{+260}_{-110}{}^{c}$ &                    &                  \\ 
NGC1333         &                             &               &                  & $344\pm173{}^{h}$         &                 &   &  \\
L134N(S)        &                             &               &                  &                           & $530^{+570}_{-180}{}^{c}$ &                    &                  \\
L1689N          &                             &               &                  &                           & $810^{+600}_{-250}{}^{c}$ &                    &                  \\
Cha-MMS1        &                             & $135{}^{i}$   &                  &                           &                 &   &     \\
\hline\noalign{\smallskip}
IRAS 17233–3606 & $276\pm62$                  &               &                  &                           &                 &   & HII region    \\ 
DR~21           &                             & $159\pm31$    & $243\pm66$       &                           &                 &   & HII region    \\
Orion~KL        &                             & $159\pm40$    & $234\pm47$       &                           &                 &   &     \\
Orion~Bar       &                             & $338\pm311$   & $361\pm141$      &                           &                 &   & PDR    \\
\hline
\end{tabular}
\tablebib{
${}^{a}$ \citet{Hilyblant13a} - the range reflects spatial variations within the core,
${}^{b}$ \citet{Hilyblant13b},
${}^{c}$ \citet{Gerin09} based on ${}^{15}$NH$_2$D observations at the IRAM~30m,
${}^{d}$ \citet{Bizzocchi13}, revised from the previously published value of $446\pm71$ \citep{Bizzocchi10},
${}^{e}$ \citet{Milam12},
${}^{f}$ \citet{Ikeda02}, 
${}^{g}$ \citet{Daniel13},
${}^{h}$ \citet{Lis10} based on observations of ${}^{15}$NH$_3$ inversion lines with the Green Bank Telescope,
${}^{i}$ reported in \citet{Milam12} based on \citet{Tennekes06},
${}^{j}$ \citet{Adande12}
}
\label{tab:prevnew}
\end{table*} 
}

\def\placetablesourceprops{
\begin{table*}[htb]
\caption{Coordinates and properties of the observed sources.}
\centering
\begin{tabular}{l@{\extracolsep{2pt}}ccccccccc}
\hline \hline\noalign{\smallskip}
Source           & RA          & Dec                                & ${v}_\mathrm{lsr}$ & Class & $d$    & $L_\mathrm{bol}$ & $M_\mathrm{env}$  \\ 
                 & [h m s]     & [${}^{\circ}~{\arcmin}~{\arcsec}$] & [km~s$^{-1}$]      &       & [pc]   & [L$_{\odot}$]    & [M$_{\odot}$]     \\
\hline
\noalign{\smallskip}
OMC-3~MMS6       & 05:35:23.5\phantom{0}${}^{a}$ & $-$05:01:32.0${}^{a}$           &           10.0 & intermed.-mass (Class 0) & 450 & 60\phantom{.0} & 36\phantom{.00} \\ 
Elias~29         & 16:27:09.4\phantom{0}${}^{b}$ & $-$24:37:18.6${}^{b}$           & \phantom{0}5.0 & Class I                  & 125 & 14.1 & \phantom{0}0.04 \\ 
IRAS~16293A      & 16:32:22.87${}^{c}$           & $-$24:28:36.4${}^{c}$           & \phantom{0}3.0 & Class 0                  & 120 & 27\phantom{.0} & \phantom{0}5.4\phantom{0}\\ 
R~CrA~IRS7B      & 19:01:56.4\phantom{0}${}^{d}$ & $-$36:57:28.3${}^{d}$           & \phantom{0}5.6 & transitionary Class 0/I  & 130 & \phantom{0}4.7 & \phantom{0}2.2\phantom{0}\\ 
\hline
\end{tabular}
\tablebib{
${}^{a}$ \citet{Johnstone03}, ${}^{b}$ \citet{Jorgensen08}, ${}^{c}$ \citet{Jorgensen11}, ${}^{d}$ \citet{Lindberg12}
}
\label{tab:sourceprops}
\end{table*} 
}

\def\placetableMoldata{
\begin{table}
\caption{Molecular data of the HCN and HNC isotopologues observed with APEX. HCN isotopologue data are from the CDMS catalog \citep{Mueller01,Mueller05} and HNC isotopologues are from the JPL catalogue \citep{Pickett98}, all obtained through www.splatalogue.net.}
\begin{center}
\begin{tabular}{l l l l l }
\hline 
\hline
Species & Transition      & Frequency & $E_\mathrm{up}$ &$\log(A_{\mathrm{ul}})$   \\ 
        &                 & [GHz]     & [K]             &[s$^{-1}$]          \\ 
\hline
\noalign{\smallskip}
\multicolumn{5}{c}{APEX-1}\\
\hline
\noalign{\smallskip}
D$^{13}$CN        & J = 3--2           & 213.51987 & 20.49 & -3.36216 \\ 
DCN               & J = 3--2           & 217.23854 & 20.85 & -3.33964 \\ 
H$^{13}$C$^{15}$N & J = 3--2           & 251.17523 & 24.11 & -3.15217 \\ 
HC$^{15}$N        & J = 3--2           & 258.15700 & 24.78 & -3.11641 \\ 
H$^{13}$CN        & J = 3--2           & 259.01180\tablefootmark{$\star$} & 24.86 & -3.11216 \\ 
HN$^{13}$C        & J = 3--2           & 261.26331 & 25.08 & -3.18837 \\
HCN               & J = 3--2           & 265.88643 & 25.52 & -3.07662 \\ 
H$^{15}$NC        & J = 3--2           & 266.58780 & 25.59 & -3.16196 \\ 
HNC               & J = 3--2           & 271.98114 & 26.11 & -2.94523 \\ 
DCN               & J = 4--3           & 289.64492 & 34.75 & -2.94907 \\ 
\hline
\noalign{\smallskip}
\multicolumn{5}{c}{APEX-2}\\
\hline
\noalign{\smallskip}
HC$^{15}$N        & J = 4--3           & 344.20011 & 41.30 & -2.72583 \\ 
H$^{13}$CN        & J = 4--3           & 345.33977\tablefootmark{$\star$} & 41.43 & -2.72161 \\ 
HN$^{13}$C        & J = 4--3           & 348.34027 & 41.80 & -2.79774 \\ 
DCN               & J = 5--4           & 362.04575 & 52.13 & -2.64860 \\ 
HNC               & J = 4--3           & 362.63030 & 43.51 & -2.55463 \\ 
\hline
\end{tabular}
\end{center}
\label{tab:moldata}
\tablefoot{\tablefoottext{$\star$} \mbox{Hyperfine structure only partially resolved.}}
\end{table}
}

\def\placetableObsRes{
\begin{table*}
\caption{Peak temperatures T$_\mathrm{peak}$, integrated intensities $\int{T_\mathrm{mb} dv}$, line widths $dv$, and central peak position $v$ of the HCN and HNC isotopologues. The integrated intensites for IRAS~16293A were obtained from integration over the line, while Gaussian fitting was used for the two other sources. Also listed are the obtained noise level (T$_\mathrm{rms}$), error on the integrated intensity $\sigma_\mathrm{rms}$, and the total error $\sigma_\mathrm{tot}$ (combined error of the integrated intensity and the calibration).}
\centering
\begin{tabular}{l@{\extracolsep{2pt}}cccccccc}
\hline \hline
\noalign{\smallskip}
Transition & T$_\mathrm{peak}$ & $\int{T_\mathrm{mb} dv}$ & $dv$ & $v$ & T$_\mathrm{rms}$ & $\sigma_\mathrm{rms}$ & $\sigma_\mathrm{tot}$\\
           & [K]               & [K~km~s$^{-1}$]          & [km~s$^{-1}$] & [km~s$^{-1}$] & [K] & [K~km~s$^{-1}$] & [K~km~s$^{-1}$]\\
\hline
\noalign{\smallskip}
\multicolumn{8}{c}{IRAS~16239A} \\
\hline\noalign{\smallskip}
HCN(3-2)                   & 9.66 &           27.52 & $6.81\pm0.02$ & $3.72\pm0.01$ & 0.04 & 0.03 & 2.3 \\ 
H${}^{13}$CN(3-2)          & 0.68 & \phantom{0}3.24 & $5.70\pm0.08$ & $3.75\pm0.03$ & 0.03 & 0.02 & 0.3 \\
HC${}^{15}$N(3-2)          & 0.27 & \phantom{0}1.37 & $6.2\pm0.2$   & $3.71\pm0.09$ & 0.04 & 0.03 & 0.1 \\
H${}^{13}$CN(4-3)          & 1.08 & \phantom{0}5.56 & $6.11\pm0.09$ & $4.28\pm0.04$ & 0.07 & 0.04 & 0.8 \\
HC${}^{15}$N(4-3)          & 0.44 & \phantom{0}2.02 & $6.4\pm0.2$   & $3.11\pm0.09$ & 0.06 & 0.04 & 0.3 \\
DCN(3-2)                   & 0.87 & \phantom{0}2.77 & $3.6\pm0.1$   & $3.87\pm0.04$   & 0.05 & 0.03 & 0.2 \\
DCN(4-3)                   & 0.76 & \phantom{0}2.45 & $3.74\pm0.07$ & $3.92\pm0.03$   & 0.05 & 0.03 & 0.4 \\
DCN(5-4)                   & 0.80 & \phantom{0}3.04 & $4.47\pm0.14$ & $3.86\pm0.06$   & 0.09 & 0.05 & 0.4 \\
D${}^{13}$CN(3-2)          & -    & -     & -             & -               & 0.02 & - & - \\
HNC(3-2)                   & 6.97 &           12.93 & $2.934\pm0.002$ & $3.959\pm0.001$ & 0.02 & 0.009 & 1.1 \\
HNC(4-3)                   & 5.57 &           11.00 & $3.14\pm0.02$   & $3.671\pm0.001$ & 0.11 & 0.05 & 0.9 \\
HN${}^{13}$C(3-2)          & 0.40 & \phantom{0}0.77 & $3.0\pm0.3$     & $3.77\pm0.09$   & 0.05 & 0.03 & 0.1 \\
H${}^{15}$NC(3-2)          & 0.09 & \phantom{0}0.22 & $2.5\pm0.4$     & $4.1\pm0.2$     & 0.03 & 0.01 & 0.03 \\
\hline
\noalign{\smallskip}
\multicolumn{8}{c}{R~CrA~IRS7B} \\
\hline\noalign{\smallskip}
HCN(3-2)          & 6.59 &           $29.48\pm0.03$ & $4.202\pm0.005$ & $5.987\pm0.002$ & 0.02 & 0.01 & 2.4 \\
H${}^{13}$CN(3-2) & 0.90 & $\phantom{0}1.87\pm0.03$ & $1.96\pm0.03$ & $5.92\pm0.01$ & 0.04 & 0.02 & 0.2 \\
HC${}^{15}$N(3-2) & 0.24 & $\phantom{0}0.45\pm0.02$ & $1.79\pm0.11$ & $5.92\pm0.05$ & 0.04 & 0.02 & 0.04 \\
H${}^{13}$CN(4-3) & 0.46 & $\phantom{0}0.92\pm0.02$ & $1.87\pm0.05$ & $6.05\pm0.02$ & 0.04 & 0.01 & 0.1 \\
HNC(3-2)          & 6.35 &           $14.3\pm0.1$ & $2.12\pm0.02$ & $5.925\pm0.008$ & 0.2 & 0.07 & 1.2 \\
HN${}^{13}$C(3-2) & 0.45 & $\phantom{0}0.75\pm0.02$ & $1.56\pm0.05$ & $5.6\pm0.02$  & 0.04 & 0.02 & 0.06 \\
H${}^{15}$NC(3-2) & 0.14 & $\phantom{0}0.20\pm0.01$ & $1.40\pm0.08$ & $5.82\pm0.03$ & 0.02 & 0.006 & 0.02 \\
HN${}^{13}$C(4-3) & 0.23 & $\phantom{0}0.32\pm0.02$ & $1.32\pm0.08$ & $5.19\pm0.04$ & 0.04 & 0.01 & 0.05 \\
\hline
\noalign{\smallskip}
\multicolumn{8}{c}{OMC-3~MMS6} \\
\hline\noalign{\smallskip}
HCN(3-2)          & 3.25 & $13.10\pm0.03$ & $3.786\pm0.009$ & $11.119\pm0.002$ &  0.01 & 0.01 & 1.09\\ 
H${}^{13}$CN(3-2) & 0.51 & $0.53\pm0.02$ & $0.98\pm0.04$ & $11.12\pm0.01$ &  0.03 & 0.01 & 0.05 \\
HC${}^{15}$N(3-2) & 0.10 & $0.10\pm0.02$ & $1.00$ (fixed) & $11.14\pm0.07$ &  0.03 & 0.01 & 0.01 \\ 
H${}^{13}$CN(4-3) & 0.18 & $0.19\pm0.04$ & $1.00$ (fixed) & $11.09\pm0.10$ &  0.06 & 0.02 & 0.03 \\ 
HNC(3-2)          & 5.55 & $6.86\pm0.09$ & $1.16\pm0.02$ & $11.177\pm0.007$ &  0.19 & 0.06 & 0.57 \\
HN${}^{13}$C(3-2) & 0.46 & $0.36\pm0.01$ & $0.74\pm0.03$ & $10.85\pm0.01$ &  0.03 & 0.008 & 0.03\\
H${}^{15}$NC(3-2) & 0.09 & $0.054\pm0.004$ & $0.54\pm0.04$ & $11.07\pm0.02$ &  0.01 & 0.003 & 0.005 \\
HN${}^{13}$C(4-3) & -    & -             & -               & -              & 0.01     & -  & - \\
\hline
\end{tabular}
\label{tab:intensities}
\end{table*} 
}

\def\placetableratios{
\begin{table}
\caption{${}^{14}$N/${}^{15}$N ratios derived from APEX observations of HCN and HNC isotopologues using a standard ${}^{12}$C/${}^{13}$C ratio of 69, except for the H${}^{13}$CN/HC${}^{15}$N(4-3) ratio, where the data are taken from the survey by \citet{Watanabe12}.  } 
\centering
\begin{tabular}{cccc}
\hline \hline
\noalign{\smallskip}
Source & $\frac{\mathrm{H}^{13}\mathrm{CN(3-2)}}{\mathrm{HC}^{15}\mathrm{N(3-2)}}$ & $\frac{\mathrm{HN}^{13}\mathrm{C(3-2)}}{\mathrm{H}^{15}\mathrm{NC(3-2)}}$ & $\frac{\mathrm{H}^{13}\mathrm{CN(4-3)}}{\mathrm{HC}^{15}\mathrm{N(4-3)}}$ \\
\noalign{\smallskip}
\hline
\noalign{\smallskip}
IRAS~16293A  & $163\pm20$ & $242\pm32$ & $190\pm38$ \\ 
R~CrA~IRS7B  & $287\pm36$ & $259\pm34$ & $(285\pm139)$ \\ 
OMC3~MMS6    & $366\pm86$ & $460\pm65$ &  -  \\ 
\hline
\end{tabular}
\label{tab:ratios}
\end{table} 
}

\def\placetablecaux{
\begin{table}
\caption{Integrated intensities from the surveys by \citet{Caux11} and \citet{Blake94,vanDishoeck95}.} 
\centering
\begin{tabular}{ccc}
\hline \hline\noalign{\smallskip}
Transition & $\int T dv$ & Telescope\\
           & [K km/s]    &  \\ 
\hline\noalign{\smallskip}
H${}^{13}$CN (1$_2$-0$_1$) & $1.04\pm0.12$ & IRAM \\ 
H${}^{13}$CN (1$_1$-0$_1$) & $0.75\pm0.09$ & IRAM \\ 
H${}^{13}$CN (1$_0$-0$_1$) & $0.25\pm0.03$ & IRAM \\ 
H${}^{13}$CN (3-2)         & $8.16\pm1.39$ & IRAM \\ 
H${}^{13}$CN (4-3)         & $6.90\pm1.24$ & JCMT \\ 
HC${}^{15}$N (1-0)         & $0.51\pm0.06$ & IRAM \\ 
HC${}^{15}$N (2-1)         & $1.88\pm0.32$ & IRAM \\ 
HC${}^{15}$N (3-2)         & $2.57\pm0.45$ & IRAM \\ 
HC${}^{15}$N (4-3)         & $1.49\pm0.27$ & JCMT \\ 
\hline\noalign{\smallskip}
H${}^{13}$CN (3-2)         & $3.45$ & JCMT \\ 
H${}^{13}$CN (4-3)         & $8.13$ & CSO \\ 
HC${}^{15}$N (3-2)         & $1.82$ & JCMT \\
HN${}^{13}$C (3-2)         & $1.33$ & JCMT \\ 
\hline
\end{tabular}
\label{tab:caux}
\end{table} 
}

\def\placetableSolarSyst{
\begin{table*}[!Htb]
\caption{Measurements of the ${}^{14}$N/${}^{15}$N ratio in solar system materials from the literature. (B) stands for ``bulk'' measurements and (HS) for ``hotspot''. Boldface indicates the unit(s) reported in the original publication (if none of the values is in boldface, the original publication reports ${}^{15}$N/${}^{14}$N). Values marked with a star were used for Fig.~\ref{fig:nitrogen_ratios}.} 
\centering
\begin{tabular}{lllll}
\hline \hline
\noalign{\smallskip}
Object & ${}^{14}$N/${}^{15}$N & $\delta {}^{15}$N & Ref. & Remarks\\
       &                       &  [\permil]        &              &        \\ 
\hline
\noalign{\smallskip}
Sun                & $441\pm6$            & $\mathbf{-383\pm8}$       & ${}^{a}$, ${}^{\star}$ & Genesis \\
Lunar soils        & $\geq 358$           & $\mathbf{\leq -240}$      & ${}^{b}$               & trapped solar wind \\
Meteorite (CAI)    & $424\pm3$            & $-359\pm6$       & ${}^{c}$, ${}^{\star}$ & Isheyevo, osbornite (TiN)\\
Jupiter            & $526$                & $-423$           & ${}^{d}$               & ISO, NH$_3$, ${}^{15}$N/${}^{14}$N = $1.9_{-1.0}^{+0.9}\times10^{-3}$ \\
Jupiter            & $435\pm57$           & $-375\pm82$      & ${}^{e}$, ${}^{\star}$ & Galileo, NH$_3$ \\  
Jupiter            & $\mathbf{448\pm62}$  & $-393\pm84$      & ${}^{f}$               & Cassini-CIRS, NH$_3$ \\ 
Jupiter            & $450\pm106$          & $-396\pm142$     & ${}^{g}$               & Cassini-CIRS, NH$_3$ \\ 
Saturn             & $\geq 357$           & $\leq -238$      & ${}^{h}$               & ground-based, NH$_3$ \\
Titan              & $\mathbf{183\pm5}$   & $486\pm41$       & ${}^{i}$               & Huygens, N$_2$, secondary origin\\
\hline\noalign{\smallskip}
Earth (atmo.)      & $\mathbf{272}\phantom{\pm00}$ & 0   & ${}^{j}$, ${}^{\star}$ & \\
Venus (atmo.)      & $272\pm54$           & $0\pm200$        & ${}^{k}$, ${}^{\star}$ & \\ 
Mars (rock)        & $280\phantom{\pm00}$ & $\mathbf{-30}$   & ${}^{l}$, ${}^{\star}$ & ALH84001 meteorite \\ 
Mars (atmo.)       & $\mathbf{173\pm11}$  & $\mathbf{572\pm82}$ & ${}^{m}$               & Curiosity, secondary origin\\
Comets             & $\mathbf{130}$ \textbf{to} $\mathbf{220}$       & $230$ to $1100$  & ${}^{n}$, ${}^{\star}$ & approx. range, CN \\   
Comet              & $\mathbf{139\pm26}$  & $957\pm366$      & ${}^{o}$               & 17P/Holmes, HCN\\ 
Comet              & $\mathbf{165\pm40}$  & $648\pm400$      & ${}^{o}$               & 17P/Holmes, CN\\
Comet              & $\mathbf{139\pm38}$  & $957\pm535$      & ${}^{p}$               & Comet C/2012 S1 (ISON), NH$_3$\\
Comets             & $\mathbf{127~(80-190)}$ & $1142$        & ${}^{q}$               & avg. spectrum (12 comets), NH$_3$\\   
Comet (B)          & $181$ to $272$       & $\mathbf{0}$ \textbf{to} $\mathbf{500}$ & ${}^{r}$ & Comet 81P/Wild2, Stardust\\
Comet (HS)         & $118$ to $368$       & $\mathbf{-260}$ \textbf{to} $\mathbf{1300\pm400}$ & ${}^{r}$ & Comet 81P/Wild2, Stardust\\
IDPs (B)           & $\mathbf{180.7\pm2.3}$ \textbf{to} $\mathbf{305.1\pm3.0}$ & $\mathbf{-108\pm9}$ \textbf{to} $\mathbf{500\pm20}$ & ${}^{s}$ & IDPs ``Kipling'', ``Eliot''\\
IDPs (HS)          & $\mathbf{118\pm6}$ \textbf{to} $\mathbf{171\pm4}$ & $\mathbf{590\pm40}$ \textbf{to} $\mathbf{1300\pm120}$ & ${}^{s}$ & IDPs ``Kipling'', ``Pirandello''\\
Cluster IDPs       & $184$ to $300$       & $\mathbf{-93\pm4}$ \textbf{to} $\mathbf{480\pm40}$ & ${}^{t}$        & IDPs ``Porky'', ``Dragonfly''\\ 
Meteorites (B)     & $192$ to $237$       & $\mathbf{146}$ \textbf{to} $\mathbf{415}$ & ${}^{u}$, ${}^{\star}$ & \\ 
Meteorites (HS)    & $\phantom{0}65$ to $193$ & $\mathbf{410\pm130}$ \textbf{to} $\mathbf{3200\pm700}$  & ${}^{u}$               & \\ 
Meteorites         & $192$ to $291$       & $\mathbf{-65.6\pm2.9}$ \textbf{to} $\mathbf{415.3\pm1.6}$ & ${}^{v}$, ${}^{\star}$ & insoluble organic matter (IOM) \\
Meteorite (HS)     & $\phantom{0}50$      & $\mathbf{4450}$  & ${}^{w}$, ${}^{\star}$ & Isheyevo clast\\
Meteorite (HS)     & $\phantom{0}46$      & $\mathbf{4900\pm300}$ & ${}^{x}$ & Isheyevo clast\\
\hline
\end{tabular}
\label{tab:solarsystem}
\tablebib{
${}^{a}$ \citet{Marty11}, 
${}^{b}$ \citet{Hashizume00}, 
${}^{c}$ \citet{Meibom07},
${}^{d}$ \citet{Fouchet00}, 
${}^{e}$ \citet{Owen01}, 
${}^{f}$ \citet{Abbas04}, 
${}^{g}$ \citet{Fouchet04},
${}^{h}$ \citet{Fletcher14},
${}^{i}$ \citet{Niemann05}
${}^{j}$ \citet{Junk58,Coplen92,Bohlke05},
${}^{k}$ \citet{Hoffman79},
${}^{l}$ \citet{Mathew01}, 
${}^{m}$ \citet{Wong13},
${}^{n}$ \citet{Manfroid09,Jehin09,Mumma11} and references therein,
${}^{o}$ \citet{Bockelee08},
${}^{p}$ \citet{Shinnaka14},
${}^{q}$ \citet{Rousselot14}, 
${}^{r}$ \citet{McKeegan06},
${}^{s}$ \citet{Floss06},
${}^{t}$ \citet{Messenger00},
${}^{u}$ \citet{Busemann06},
${}^{v}$ \citet{Alexander07},
${}^{w}$ \citet{Bonal10},  
${}^{x}$ \citet{Briani09}
}
\end{table*} 
}

\def\placetableobsdetail{
\begin{table*}
\caption{Observational details.} 
\centering
\begin{tabular}{llccl}
\hline \hline
Transition(s) & Date & \# scans & time (on) & Remarks \\
              &      &          & [min]     &         \\
\hline 
\noalign{\smallskip}
\multicolumn{5}{c}{IRAS~16293A}\\
\hline
\noalign{\smallskip}
H${}^{13}$CN(3-2), HC${}^{15}$N(3-2)                                     & 2012-04-01 & \phantom{0}67 & 20 & \\ 
H${}^{13}$CN(4-3), HC${}^{15}$N(4-3)                                     & 2012-06-04 & \phantom{0}41 & 12 & \\ 
H${}^{13}$C${}^{15}$N(3-2)                                               & 2012-06-13 & 150 & 44 & USB (30 scans), LSB (120 scans)\\ 
                                                                         &            &     &    & LSB suffers from strong spur \\
DCN(3-2)                                                                 & 2012-06-08 & \phantom{0}29 & \phantom{0}9 & scans 47-48 ignored (LO locked out)\\
DCN(4-3)                                                                 & 2012-06-07 & \phantom{0}30 & \phantom{0}9 & \\
DCN(5-4), HNC(4-3)                                                       & 2012-06-04 & \phantom{0}40 & 12 & baseline ripples\\
D${}^{13}$CN(3-2)                                                        & 2012-06-08 &  120 & 35 & scans 541-542 ignored (LO locked out)\\
                                                                         & 2012-06-09 & \phantom{0}41 & 12 & scans 215-218 ignored (LO locked out)\\
                                                                         & 2012-06-10 & \phantom{0}50 & 15 \\
                                                                         & 2012-06-11 & \phantom{0}60 & 18 \\
HN${}^{13}$C(3-2), H${}^{13}$CN(3-2), HC${}^{15}$N(3-2)                  & 2013-08-20 & \phantom{0}36 & 21  & H${}^{13}$CN and HC${}^{15}$N intensities higher than \\
                                                                         &            &               &     & those observed on 2012-04-01\\
H${}^{15}$NC(3-2), HCN(3-2)                                              & 2013-08-19 & \phantom{0}36 & 21 \\
\hline
\noalign{\smallskip}
\multicolumn{5}{c}{R~CrA~IRS~7B}\\
\hline
\noalign{\smallskip}
HN${}^{13}$C(3-2), H${}^{13}$CN(3-2), HC${}^{15}$N(3-2)                  & 2012-08-16 & \phantom{0}30 & \phantom{0}9 & \\
                                                                         & 2012-09-27 & \phantom{0}50 & 15 \\
HN${}^{13}$C(4-3), H${}^{13}$CN(4-3)                                     & 2012-08-22 & \phantom{0}50 & 15 \\
                                                                         & 2012-09-25 & \phantom{0}30 & \phantom{0}9 & \\
H${}^{15}$N(3-2)C, HCN(3-2)                                              & 2012-08-16 & 108 & 32 \\
                                                                         & 2012-09-27 & 140 & 41 \\
                                                                         & 2012-09-29 & \phantom{0}90 & 26 \\
                                                                         & 2012-09-30 & \phantom{0}40 & 12 & intensities \& line profiles off, data discarded\\
                                                                         & 2012-11-20 & \phantom{0}30 & \phantom{0}9 & \\
                                                                         & 2012-11-21 & \phantom{0}40 & 12 \\
                                                                         & 2012-11-23 & 100 & 29 \\
                                                                         & 2012-11-24 & \phantom{0}10 & \phantom{0}3 & \\
                                                                         & 2012-11-28 & \phantom{0}50 & 15 \\
HNC(3-2)C                                                                & 2012-09-27 & \phantom{0}20 & \phantom{0}6 \\
\hline
\noalign{\smallskip}
\multicolumn{5}{c}{OMC3~MMS6} \\
\hline
\noalign{\smallskip}
HN${}^{13}$C(3-2), H${}^{13}$CN(3-2), HC${}^{15}$N(3-2)                  & 2012-08-16 & \phantom{0}28 & \phantom{0}8 & \\
                                                                         & 2012-11-22 & 120 & 35 \\
HN${}^{13}$C(4-3), H${}^{13}$CN(4-3)                                     & 2012-09-26 & \phantom{0}30 & \phantom{0}9 \\
H${}^{15}$NC(3-2), HCN(3-2)                                              & 2012-08-16 & \phantom{0}70 & 21 \\
                                                                         & 2012-09-25 & 170 & 50 \\
                                                                         & 2012-11-24 & \phantom{0}90 & 26 \\
                                                                         & 2012-11-25 & \phantom{0}70 & 21 \\
                                                                         & 2012-11-26 & 190 & 56 \\
HNC(3-2)C                                                                & 2012-09-27 & \phantom{0}10 & \phantom{0}3 \\
\hline
\noalign{\smallskip}
\multicolumn{5}{c}{Elias~29} \\
\hline
\noalign{\smallskip}
HN${}^{13}$C(3-2), H${}^{13}$CN(3-2), HC${}^{15}$N(3-2)                  & 2012-08-16 & \phantom{0}30 & \phantom{0}9 & \\ 
HN${}^{13}$C(4-3), H${}^{13}$CN(4-3)                                     & 2012-09-30 & \phantom{0}80 & 24 & \\ 
H${}^{15}$N(3-2)C, HCN(3-2)                                              & 2012-08-16 & \phantom{0}50 & 15 & intensity on the two days is very different\\
                                                                         & 2012-10-01 & \phantom{0}76 & 23 & (factor of $\sim 1.5$), data not used \\
HNC(3-2)C                                                                & 2012-09-30 & 160 & 47 & \\ 
                                                                         & 2012-10-01 & \phantom{0}50 & 15 & \\ 
\hline
\end{tabular}
\label{tab:obsdetails}
\end{table*} 
}


\abstract
   {The terrestrial planets, comets, and meteorites are significantly enriched in $^{15}$N compared to the Sun and Jupiter. While the solar and jovian nitrogen isotope ratio is believed to represent the composition of the protosolar nebula, a still unidentified process has caused $^{15}$N-enrichment in the solids. Several mechanisms have been proposed to explain the variations, including chemical fractionation. However, observational results that constrain the fractionation models are scarce. While there is evidence of $^{15}$N-enrichment in prestellar cores, it is unclear how the signature evolves into the protostellar phases.}
   {The aim of this study is to measure the $^{14}$N/$^{15}$N ratio around three nearby, embedded low- to intermediate-mass protostars.}
   {Isotopologues of HCN and HNC were used to probe the $^{14}$N/$^{15}$N ratio. A selection of J=3-2 and 4-3 transitions of H$^{13}$CN, HC$^{15}$N, HN$^{13}$C, and H$^{15}$NC was observed with the Atacama Pathfinder EXperiment telescope (APEX). The $^{14}$N/$^{15}$N ratios were derived from the integrated intensities assuming a standard $^{12}$C/$^{13}$C ratio. The assumption of optically thin emission was verified using radiative transfer modeling and hyperfine structure fitting.} 
   {Two sources, IRAS~16293A and R~CrA~IRS7B, show $^{15}$N-enrichment by a factor of $\sim1.5-2.5$ in both HCN and HNC with respect to the solar composition. IRAS~16293A falls in the range of typical prestellar core values. Solar composition cannot be excluded for the third source, OMC-3~MMS6. Furthermore, there are indications of a trend toward increasing $^{14}$N/$^{15}$N ratios with increasing outer envelope temperature.} 
   {The enhanced $^{15}$N abundances in HCN and HNC found in two Class~0 sources ($^{14}\mathrm{N}/^{15}\mathrm{N} \sim 160-290$) and the tentative trend toward a temperature-dependent $^{14}$N/$^{15}$N ratio are consistent with the chemical fractionation scenario, but $^{14}$N/$^{15}$N ratios from additional tracers are indispensable for testing the models. Spatially resolved observations are needed to distinguish between chemical fractionation and isotope-selective photochemistry.}


\maketitle

\section{Introduction}
\renewcommand{\thefootnote}{\fnsymbol{footnote}}

The largest isotopic anomalies in the solar system are found in highly volatile elements like hydrogen, carbon, oxygen, and nitrogen \citep[e.g.,][]{Birck04,Anders89}. 
Thereof, the most extreme variations among solar system objects exist in the deuterium-to-hydrogen ratio, D/H, and the ratio of the two stable nitrogen isotopes, $^{14}$N/$^{15}$N\footnote{In cosmochemistry, nitrogen isotope ratios are generally reported in per mil deviations from the terrestrial atmospheric standard, denoted $\delta {}^{15}\mathrm{N} = \left[\left(^{15}\mathrm{N}/^{14}\mathrm{N}\right)_\mathrm{sample}/\left(^{15}\mathrm{N}/^{14}\mathrm{N}\right)_\mathrm{air} -1 \right]\times 10^3$ where ${}^{15}\mathrm{N}/^{14}\mathrm{N}_\mathrm{air} = 3.676\times10^{-3}$.}. 
Recent results from the Genesis mission \citep{Marty11}, which sampled the solar wind, have demonstrated that there is a large discrepancy in the nitrogen isotopic composition between the solar system solids and the gas reservoir: The $^{14}$N/$^{15}$N ratios of the terrestrial planets, comets \citep[e.g.,][]{Hutsemekers05,McKeegan06,Bockelee08,Manfroid09,Jehin09,Mumma11,Shinnaka14}, meteorites \citep[e.g.,][]{Kerridge93,Briani09,Bonal10}, and interplanetary dust particles \citep[IDPs,][]{Messenger00,Floss06} are significantly enhanced compared to the Sun \citep[$^{14}$N/$^{15}$N $= 441\pm6$,][]{Marty11} and Jupiter \citep[$435\pm57$,][]{Owen01,Fouchet04,Abbas04}. The derived ratios for the Earth's atmosphere \citep[272, e.g.,][]{Junk58,Mariotti83}, Venus \citep{Hoffman79}, and the interior of Mars \citep{Mathew01} are very similar and also lie within the range spanned by most primitive meteorites \citep[e.g., Fig.~2 from][]{Meibom07}. 
The nitrogen isotopic composition of the Sun and Jupiter, on the other hand, is substantially $^{15}$N-poorer \citep[][]{Marty11,Hashizume00} and believed to be representative of the presolar nebula. This idea is supported by the results from an isotopic measurement of a high-temperature condensate -- titanium nitride (TiN), named osbornite -- in a calcium-aluminum-rich inclusion (CAI) of the ``Isheyevo'' meteorite. CAIs are believed to be the first solids that condensated from the gas of the presolar nebula. The $^{14}$N$/^{15}$N ratio measured from this osbornite \citep[$424\pm3$,][]{Meibom07} is therefore interpreted as reflecting the nitrogen isotope composition of the solar nebula at the time of CAI formation $4567.3\pm0.16~\mathrm{Myr}$ ago \citep[][]{Connelly12}. 

The aim of this paper is to study the $^{14}$N/$^{15}$N ratio around low-mass protostars to investigate whether fractionation of nitrogen isotopes occurs in the early phases of star formation and to better understand what mechanisms could be responsible for the solar system heterogeneity. 

\renewcommand{\thefootnote}{\arabic{footnote}}
\setcounter{footnote}{0}

\subsection{${}^{15}$N-enrichments in primitive solar system matter}
Meteorites, interstellar dust particles, and comets represent the most pristine solar system material. For meteorites, the nitrogen isotope composition is highly variable among the different classification groups, among meteorites belonging to the same group, and even within individual meteorites \citep[e.g.,][]{Alexander07,Bonal10}. The heaviest nitrogen isotope signature in any solar system material was measured to date in the ``Isheyevo'' chondrite \citep[$^{14}$N/$^{15}$N $\sim 50$,][]{Briani09,Bonal10}. However, the nitrogen isotope composition in Isheyevo is also extremely variable, ranging from an almost solar-like (${}^{14}\mathrm{N}/{}^{15}\mathrm{N} = 394\pm12$) to a very ${}^{15}$N-heavy isotopic composition. The $^{15}$N-enrichment consists of two components, a diffuse background enrichment (${}^{14}\mathrm{N}/{}^{15}\mathrm{N}\sim 160$) and localized micron-sized zones with extremely high $^{15}$N-anomalies, so-called hotspots. Generally, the highest $^{15}$N-enrichments seem to be found in the most pristine meteoritic components \citep{Busemann06,Alexander07,Briani09,Bonal10}. The carrier of the signature is most likely carbonaceous \citep{Bonal10}, but the exact functional groups carrying the $^{15}$N-enrichment have not yet been identified.

Interplanetary dust particles (IDPs) are collected in the stratosphere of the Earth and believed to be of cometary or potentially asteroidal origin. They show similar $^{15}$N-characteristics as chondrites, with a modest diffuse enrichment and hotspots, but with generally more moderate $^{15}$N-enhancements \citep{Busemann06,Floss06}. The ``Stardust'' samples from comet 81P/Wild-2 have bulk nitrogen isotope compositions that cluster around the terrestrial composition, but moderate enhancements are also measured \citep{McKeegan06}. Hotspot values on submicrometer scale are similar to IDPs. An isotopically light component, consistent with the solar value, is also found in the Wild-2 particles, which is not the case for IDPs.

The average $^{14}$N/$^{15}$N ratio from CN spectroscopy in a sample of comets is $147.8\pm5.6$ \citep{Manfroid09} with slight differences for Jupiter family ($^{14}\mathrm{N}/^{15}\mathrm{N} = 156.8\pm12.2$, probable Kuiper belt origin) and Oort cloud ($^{14}\mathrm{N}/^{15}\mathrm{N} = 144\pm6.5$) comets. These results are consistent with the $^{14}$N/$^{15}$N ratio from HCN measurements by \citet{Bockelee08}  in the comets ``Hale-Bopp'' and ``Holmes'' \citep[see also the review by][]{Jehin09} as well as measurements from NH$_2$ in comet C/2012 S1 \citep[ISON, ${}^{14}\mathrm{N}/{}^{15}\mathrm{N} = 139\pm38$,][]{Shinnaka14} and the average spectrum of 12 comets \citep[$\sim 130$,][]{Rousselot14}.

An overview of $^{14}$N/$^{15}$N ratios in solar system bodies from the literature is provided in Table~\ref{tab:solarsystem} in the appendix.

\subsection{Proposed scenarios: chemical fractionation vs. isotope selective photodissociation}

\placetableprevnew

\placetablesourceprops

A variety of processes have been proposed to explain the observed nitrogen isotope heterogeneity, including Galactic chemical evolution \citep[GCE, e.g.,][and references therein]{Adande12}, isotope-selective N$_2$ photodissociation \citep[][]{Clayton02,Lyons09,Heays14}, and chemical fractionation \citep{Terzieva00,Rodgers08b,Rodgers08a,Wirstroem12,Hilyblant13a}. 

The term ``Galactic chemical evolution'' refers to a scenario in which the observed $^{14}$N/$^{15}$N ratio variations are the result of an enrichment of the interstellar medium (ISM) with the products of stellar nucleosynthesis over time. GCE models predict the $^{14}$N/$^{15}$N ratio to vary with distance from the Galactic center as a result of different production mechanisms for $^{14}$N and $^{15}$N, the details of which are not yet fully understood \citep[see discussion and references in][]{Adande12}. Observational support comes from surveys of the nitrogen isotopic composition of the Galaxy \citep[by, e.g.,][]{Linke77,Wannier81,Guesten85,Dahmen95,Wilson99,Adande12}. 
Most of these studies found a radial gradient in the $^{14}$N/$^{15}$N ratio of the Galaxy, arguing for Galactic evolution as the main cause for deviations in the isotopic composition of different star-forming regions. However, significant source-to-source variations within individual clouds \citep[e.g.,][]{Ikeda02} as well as evidence for spatial variations in prestellar cores \citep{Hilyblant13a} hint at local processes being more important for the nitrogen isotope composition of the protosolar disk than Galactic inheritance. 

\citet{Liang07b} demonstrated that isotopically selective photodissociation of N$_2$ plays an important role in the explanation of the nitrogen isotope anomaly inferred in the upper atmosphere of Titan from HCN observations \citep{Niemann05}. The radiation field component corresponding to the dissociation energy of $^{14}$N$_2$ cannot penetrate as deep into the atmosphere as the one for $^{14}$N$^{15}$N, because of $^{14}$N$_2$ being more abundant than $^{14}$N$^{15}$N, an effect which is referred to as photochemical self-shielding. To what extent N$_2$ self-shielding could have affected the nitrogen isotope composition of the protosolar nebula is not yet clear, but is currently being investigated \citep[e.g.,][]{Lyons09,Bertin13}. \citet{Li13} and \citet{Heays14} have recently presented photodissociation rates for $^{14}$N$_2$ and $^{14}$N$^{15}$N . In addition, \citet{Heays14} have incorporated these rates into chemical models with conditions simulating an interstellar cloud and a protoplanetary disk with grain growth. Their results show $^{15}$N-enhancements by up to a factor of 10 in HCN in the protoplanetary disk model. Based on these results they argue that the observed nitrogen fractionation in protostellar environments could be caused by a combination of chemical and photolytic reactions. 

As for chemical fractionation, the astrochemical gas-grain models by \citet{Rodgers08b} predict a considerable $^{15}$N-enhancement for different nitrogen-bearing molecules in the gas phase of dense cloud cores through isotope exchange reactions in low-temperature chemistry. Their results indicate that the fractionation mechanisms for nitriles, i.e., molecules bearing the nitrile functional group (-CN), and nitrogen hydrides (-NH) are different. The slow route ($\sim 10^6~\mathrm{yrs}$) through the dissociative ionization of N$_2$, leading to N$^{+}$, is responsible for the fractionation in nitrogen hydrides, including ammonia (NH$_3$) and diazenylium (N$_2$H$^{+}$). The more rapid route ($\sim 10^5~\mathrm{yrs}$) starts from atomic nitrogen and drives the $^{15}$N-enrichment of (iso-)nitriles and associated species such as cyanide (CN), hydrogen cyanide (HCN), and hydrogen isocyanide (HNC). The largest $^{15}$N-enhancement in this chemical fractionation model is predicted for nitrile species. With substantial $^{15}$N-enrichments predicted by both chemical fractionation and photodissociation models, species like HCN should therefore be particularly good tracers of a potential isotope fractionation in nitrogen. Refined chemical fractionation models by \citet{Wirstroem12} include the dependence of the ion-molecule reactions on the H$_2$ spin-state and show that a large $^{15}$N-enhancement is not necessarily expected to correlate with an extreme D-enrichment, depending on the carriers of the two signatures. The lack of a correlation between the $^{15}$N and D anomalies measured in meteorites had previously often been invoked as an argument against an inheritance of the isotopic signature measured in primitive solar system matter from local cloud or disk processes in early phases of solar system formation \citep[e.g.,][]{Aleon10}. 

Observations of the $^{14}$N/$^{15}$N ratio in individual low-mass pre- and protostellar sources in the literature are scarce and have so far led to inconclusive results about whether or not fractionation of nitrogen isotopes occurs in pre- and protostellar environments. Moreover, it still an open question whether chemical fractionation or isotope-selective photodissociation could explain the observed solar system variations. Results from HCN observations of three dense cores by \citet{Ikeda02} show large variations between the sources and the result for L1521E ($^{14}\mathrm{N}/^{15}\mathrm{N} = 151\pm16$) is consistent with the chemical fractionation scenario. \citet{Gerin09} on the other hand derive $^{14}$N/$^{15}$N ratios of $\sim 360-810$ from $^{15}$NH$_2$D in several interstellar clouds that are consistent with the solar composition and thus argue against nitrogen isotope fractionation. Follow-up observations by \citet{Lis10} using measurements of $^{15}$NH$_3$ in Barnard~1 and NGC~1333 yield ratios that fall in between the solar and terrestrial values. \citet{Bizzocchi10} reported a ${}^{14}\mathrm{N}/{}^{15}\mathrm{N}$ ratio of $446\pm71$ from N$^{15}$NH$^+$ observations toward the prestellar core L1544, which was recently revised to be as high as $1000\pm200$ based on new observations of ${}^{15}$NNH$^+$  \citep{Bizzocchi13}, implying a depletion of ${}^{15}$N in N$_2$H$^{+}$. Results from mapping of HCN isotopologues in the two prestellar cores L183 and L1544 by \citet{Hilyblant13a} show an overall enhancement of $^{15}$N as well as some spatial variations within each source, strongly favoring a fractionation scenario. For L1544, \citet{Hilyblant13a} find an enhancement of $^{15}$N in HCN ($140 < {}^{14}\mathrm{N}/^{15}\mathrm{N} < 360 $), but there are no indications for isotope fractionation in CN \citep[${}^{14}\mathrm{N}/^{15}\mathrm{N} = 500\pm75$,][]{Hilyblant13b}. These differences likely arise from the variety of molecular species and isotopologues used to probe the nitrogen isotope ratio in different sources. For example, the low $^{14}$N/$^{15}$N ratios inferred from nitriles like HCN seem to support the chemical fractionation scenario, while $^{14}$N/$^{15}$N ratios derived from nitrogen hydrides, like NH$_3$ and N$_2$H$^{+}$, show no evidence for nitrogen isotope fractionation or even hint at a $^{15}$N-depletion. \citet{Wirstroem12} and \citet{Hilyblant13b} proposed refined chemical fractionation scenarios with different fractionation paths that may be able to explain the discrepancies between the nitrogen isotope composition seen in individual molecular tracers. To constrain the fractionation models, it is thus important to measure isotope ratios from an entire set of species. However, the prestellar core L1544 and the very young Class~0 binary B1b are so far the only sources for which comprehensive observational data sets of $^{14}$N/$^{15}$N ratios from both nitriles and nitrogen hydrides exist (cf. Table~\ref{tab:prevnew}). While evidence for a $^{15}$N-enrichment in some molecules is found for L1544 \citep{Hilyblant13a}, the results on B1b by \citet{Daniel13} show no evidence for nitrogen isotope fractionation in any class of molecules. Whether this difference is caused by the more evolved status of B1b, a source-to-source variation, or any other effect, is unclear and needs to be investigated on a larger source sample. A similar data set is not yet available for other sources, in particular not for more evolved evolutionary stages.

In contrast to previous studies, which mainly focused on the prestellar core phase, we present in this paper measurements of the nitrogen isotope composition in the gas around three Class~0 YSOs from HCN and HNC isotopologues. The aim of this work is to investigate whether signatures of nitrogen isotope fractionation also occur in the embedded protostellar stages. The paper is organized as follows: Section~\ref{sec:observations} describes the observations and the data reduction procedure. The results are presented in Sect.~\ref{sec:results}. They are compared to previous results and their implications are discussed in Sect.~\ref{sec:discussion}. Conclusions are presented in Sect.~\ref{sec:conclusions}.

\section{Observations and data reduction} \label{sec:observations}
\subsection{Observations}
 
\placetableMoldata

Observations of HCN and HNC isotopologues were carried out using the Atacama Pathfinder EXperiment telescope\footnote{APEX is a collaboration between the Max-Planck-Institut f\"{u}r Radioastronomie, the European Southern Observatory, and the Onsala Space Observatory.} \citep[APEX,][]{Guesten06} in service mode. Wobbler switching with an amplitude of 150$\arcsec$ was used throughout. A detailed overview on the observations is provided in Table~\ref{tab:obsdetails} in the appendix. 

The source sample consists of three Class~0 YSOs, IRAS~16293A, R~CrA~IRS7B, and OMC-3~MMS6, and the Class~I protostar Elias~29. The coordinates and properties of the sources can be found in Table~\ref{tab:sourceprops}. The targets were selected to cover different source properties: IRAS~16293-2422 is an extremely well-studied \citep[e.g.,][]{Blake94,vanDishoeck95,Schoier02,Takakuwa07,Caux11,Jorgensen11,Pineda12}, isolated proto-binary in $\rho$ Oph with a hot corino and known deuterium fractionation \citep[e.g.,][]{Loinard00,Persson13}. Our observations were pointed at the A component, IRAS~16293A \citep{Jorgensen11}. R~CrA~IRS7B is located in the vicinity of the Herbig star R~CrA and therefore exposed to a strong external radiation field \citep{Lindberg12}. OMC-3~MMS6 is an intermediate-mass protostar \citep{Johnstone03} in the vigorous environment of the Orion star-forming complex, also exposed to enhanced irradiation \citep{Jorgensen06}. Elias~29 \citep[e.g.,][]{Boogert02} is one of the most luminous protostars in the $\rho$ Oph cloud and the most evolved source in our sample.   

IRAS~16293A was observed in H${}^{13}$CN (\mbox{$J$ = 3-2} and \mbox{4-3}), HC${}^{15}$N (\mbox{3-2} and \mbox{4-3}), H${}^{13}$C${}^{15}$N(3-2), DCN(3-2, \mbox{4-3}, and \mbox{5-4}), and D${}^{13}$CN(3-2) on 2012-04-01 and between 2012-06-04 and 2012-06-13 (project \mbox{089.F-9316A}, PI J.~J\o{}rgensen). The observations of HN$^{13}$C(3-2) and H$^{15}$NC(3-2) were carried out on 2013-08-19/20 (project \mbox{092.F-9324A}, PI S.~Wampfler). OMC-3~MMS6, Elias~29, and R~CrA~IRS7B were observed in HCN(3-2), H${}^{13}$CN (\mbox{3-2} and \mbox{4-3}), HC${}^{15}$N(3-2), HNC(3-2), HN${}^{13}$C (\mbox{3-2} and \mbox{4-3}), and H${}^{15}$NC(3-2) with APEX between 2012-08-16 and 2012-10-01 (project \mbox{090.F-9317A}, PI S.~Wampfler). The frequencies of the transitions and the molecular data are provided in Table~\ref{tab:moldata}. 

\placefigureIRASspec

The Swedish Heterodyne Facility Instrument \citep[SHeFI,][]{Vassilev08} single-sideband SIS receivers APEX-1 ($211-275$~GHz) and APEX-2 ($267-370$~GHz) were used in combination with the eXtended bandwidth Fast Fourier Transform Spectrometer (XFFTS). Two XFFTS units with an instantaneous bandwidth of 2.5~GHz each are available, which partially overlap and offer full coverage of the 4~GHz IF bandwidth of the SHeFI receivers. The large bandwidth of this backend type permits several isotopologues to be observed simultaneously, thereby reducing the calibration uncertainty from the absolute value down to the relative calibration uncertainty across the band. 
For IRAS~16293A, H${}^{13}$CN (3-2) and HC${}^{15}$N (3-2) were observed in a combined setting, as were H${}^{13}$CN (4-3) and HC${}^{15}$N (4-3). In the case of OMC-3~MMS6, Elias~29, and R~CrA-IRS7B, the observations of H${}^{13}$CN (3-2), HC${}^{15}$N (3-2), and HN${}^{13}$C (3-2) were carried out simultaneously. Moreover, H${}^{15}$NC (3-2) and HCN(3-2) were observed together. H${}^{13}$CN (4-3) and HN${}^{13}$C (4-3) were also jointly measured.

\placetableObsRes

\subsection{Data reduction}

The data reduction was carried out using the GILDAS-CLASS software package\footnote{\url{http://www.iram.fr/IRAMFR/GILDAS/}}.  
The data were calibrated to $T_\mathrm{mb}$ scale using the default forward efficiency of 0.95 \citep{Dumke10} for APEX, and beam efficiencies\footnote{values from \url{http://www.apex-telescope.org/telescope/efficiency}} of 0.75 for APEX-1 and 0.73 for APEX-2, respectively. The half-power beam width of APEX is 27$\arcsec$ at 230~GHz and 18$\arcsec$ at 345~GHz. 

Several observations were split between different observing dates (cf. Table~\ref{tab:obsdetails}). Before averaging those data sets, we checked for deviations in the intensity and line profile on strong lines present in the spectrum, and, if the signal-to-noise ratio for the individual observations permits, also on our target lines. 
The H$^{15}$NC(3-2) observations from R~CrA~IRS7B carried out on the 09-30-2012 show substantially larger deviations than the calibration uncertainty ($\sim30\%$) when compared to other days and this dataset was consequently discarded. Deviations in the line profiles between different days also occur for the detected main isotopologues lines HCN and HNC in Elias~29 (rarer isotopologues are not detected). Moreover, the line profiles are a mixture of absorption and emission, which could be an indication for emission in one of the off positions (wobbler switching mode). Therefore, we do not report integrated intensities for Elias~29. 

A potential contamination from the off positions for the other sources was assessed by re-pipelining some data sets using only one or the other off position for H$^{13}$CN and H$^{13}$CO$^+$, a stronger line in the same frequency setup. The resulting spectra, which can be found in appendix \ref{sec:obsdetails}, were then examined for differences. There is no evidence for a substantial emission from the off positions.
 
First order baselines were applied to the individual scans before averaging them. In particular, the DCN(5-4) spectrum suffers from standing waves, which cannot easily be fitted with a sinusoidal baseline, as amplitude and period change. Therefore, the first order polynomial baseline fits were applied locally whenever required. For lines that were covered by both XFFTS units (overlapping spectral range), the processed spectra from the individual parts were finally averaged after visual inspection.

\section{Results} \label{sec:results}
\subsection{Line intensities}\label{sec:intensites}
\placefigureRCrAspec

The line intensities were obtained by Gaussian fitting except for IRAS~16293A, because the line shapes of this source differ substantially from a simple Gaussian profile (cf. Fig.~\ref{fig:iras16293_spectra}). For IRAS~16293A, the intensities were derived from integrating over a velocity range around the source velocity  ($-10~\mathrm{km}~\mathrm{s}^{-1} < \varv < 16~\mathrm{km}~\mathrm{s}^{-1}$ for broad lines and $-4~\mathrm{km}~\mathrm{s}^{-1} < \varv < 12~\mathrm{km}~\mathrm{s}^{-1}$ for narrow lines). Table~\ref{tab:intensities} lists the resulting integrated line intensities, widths, and the obtained noise level (rms). The spectra from R~CrA~IRS7B and OMC-3~MMS6 are shown in Figs.~\ref{fig:rcrairs7b_spectra} and \ref{fig:omc3mms6_spectra}, respectively.

While the HCN(3-2) line has prominent outflow wings in IRAS~16293A and OMC-3~MMS6, the line width of HNC(3-2) is narrower and does not feature the broad wings at the obtained signal-to-noise ratio. 

For IRAS~16293A, H$^{13}$C$^{15}$N(3-2) and D$^{13}$CN(3-2) were targeted as well, but not detected at the achieved noise level. The H$^{13}$C$^{15}$N(3-2) was only observed for a fraction of the time intended, and the sideband was changed during the observation because of a better receiver performance in the upper sideband. Unfortunately, the upper sideband tuning suffers from a tuning instability, which went unnoticed during the observations. A reflection from the tuning instability shows up in one channel at the tuning frequency, and a reliable determination of the noise level is therefore not possible. Hence, we do not give an upper limit on the H$^{13}$C$^{15}$N(3-2) intensity.

In the Class~I source Elias~29, the only lines detected are HCN(3-2) and HNC(3-2), which are factors of 50-100 times weaker than the same lines from the Class~0 targets. Therefore, none of the rarer isotopologues are detected in Elias~29 at the achieved rms, maybe because of the much smaller envelope mass compared to the Class~0 sources in the sample. Both HCN(3-2) and HNC(3-2) have a line profile that shows emission at velocities higher than the source velocity, and absorption at the lower velocities, which could either be a real source feature or indicate that there was some molecular emission at the off position.

\placefigureOMCspec

\subsection{${}^{14}$N/${}^{15}$N isotopic ratios}\label{sec:ratios}

\placefigureRatios

Isotope ratios can generally be measured from molecular lines at sub-millimeter wavelengths using three different approaches, each with its advantages and caveats:
\begin{enumerate}
 \item \textit{Singly substituted molecules:} The ${}^{14}$N/${}^{15}$N ratio can be measured from, e.g., CN isotopologues \citep[C$^{15}$N][]{Adande12,Hilyblant13b}, N$_2$H$^+$ isotopologues \citep[$^{15}$NNH$^+$, $^{15}$NNH$^+$,][]{Womack92,Bizzocchi10,Bizzocchi13} or ammonia isotopologues \citep[$^{15}$NH$_3$,][]{Guesten85,Lis10}. The advantage of this method is that the main isotopologue is usually abundant and thus its lines are strong. On the other hand, the high abundance often implies a high optical depth, which must be accounted for using radiative transfer modeling. If the molecular transition has a resolved hyperfine structure pattern, the intensity ratios of the hyperfine transitions can also be used to infer and correct for the optical depth.  
 \item \textit{Doubly substituted molecules:} An alternative to the often optically thick main isotopologues are molecules with two atoms replaced by a less abundant isotope, like the ammonia isotopologue ${}^{15}$NH$_2$D \citep[][]{Gerin09}. However, the lines of such doubly substituted molecules are often very weak because of their low abundance. 
 \item \textit{Double isotope method:} The nitrogen isotope ratio can also be measured from two singly substituted molecules, e.g., H$^{13}$CN and HC$^{15}$N or HN$^{13}$C and H${}^{15}$NC, avoiding the main isotopologue. However, in this example the ${}^{12}$C/${}^{13}$C ratio needs to be known from independent measurement or an assumption has to be made.   
\end{enumerate}

Using the double isotope method and assuming that all lines are optically thin, which will be discussed in Sect.~\ref{sec:optdepth}, the ${}^{14}$N/${}^{15}$N ratio can be obtained from the H$^{13}$CN and HC$^{15}$N integrated intensities using the relation
\begin{center}
\begin{eqnarray}
 \frac{{}^{14}\mathrm{N}}{{}^{15}\mathrm{N}} &=& \frac{\int T dv~(\mathrm{H}{}^{13}\mathrm{CN})}{\int T dv~(\mathrm{HC}^{15}\mathrm{N})} \times \frac{{}^{12}\mathrm{C}}{{}^{13}\mathrm{C}} \label{eqn:simpleratio}
\end{eqnarray}
\end{center}
and from HNC isotopologues accordingly. For IRAS~16293A, the non-detection of D$^{13}$CN hampers a direct determination of the ${}^{12}$C/${}^{13}$C from D$^{13}$CN and DCN, but only provides a $3\sigma$ upper limit of ${}^{12}\mathrm{C}/{}^{13}\mathrm{C} \lesssim 120$. Because no ${}^{12}$C/${}^{13}$C ratio measurements for our sources are available from the literature, we assume the local standard ISM value of 69 \citep{Wilson99,Milam05}. The resulting ratios are illustrated in Fig.~\ref{fig:nitrogen_ratios}, together with solar system measurements from the literature, and listed in Table~\ref{tab:ratios}. The table also includes the ratio from the integrated intensities of H$^{13}$CN(4-3) and HC$^{15}$N(4-3) reported in \citet{Watanabe12}. The H$^{13}$CN(4-3) line may have a contribution of SO$_2$($13_{2,12}-12_{1,11}$), but given that the isotopic ratios derived from H$^{13}$CN(3-2) and (4-3) in IRAS~16293A agree well, the contamination is expected to be minor in this source. For the other sources, H$^{13}$CN(4-3) was not used in the determination of the ${}^{14}$N/${}^{15}$N ratio, just as a constraint for the radiative transfer models. 

\placetableratios

The smallest error is expected on the ratio inferred from H$^{13}$CN(3-2) and HC$^{15}$N(3-2) for all sources as well as the H$^{13}$CN(4-3) and HC$^{15}$N(4-3) for IRAS~16293A, because the lines were measured simultaneously. The  ${}^{14}$N/${}^{15}$N ratios from the HNC isotopologues are more uncertain, as the involved lines were observed in two individual spectral setups. 
For the errors, we use a combined error from the intensity measurement and the calibration uncertainty. The error on the integrated intensity is taken from the uncertainty in the fit of the area under the Gaussian profile, provided by the CLASS software, for R~CrA~IRS7B and OMC-3~MMS6, and for IRAS~16293A calculated as $\sigma_\mathrm{rms} = \sqrt{\Delta \varv \times \delta v} \times T_\mathrm{rms}$ where $\Delta \varv$ is the full width at half maximum of the line (assumed to be the same as that of a Gaussian fit to the line), $\delta v$ is the velocity resolution of the spectrum, and $T_\mathrm{rms}$ is the noise level. For the calibration error  $\sigma_\mathrm{calib}$, we used the calibration uncertainties on the integrated intensity provided in the APEX calibration report \citep{Dumke10} of 8.3\% and 14.1\% for APEX-1 and APEX-2, respectively. The overall error on a line intensity is then the combination of the two errors, $\sigma_\mathrm{tot} = \sqrt{\sigma_\mathrm{rms}^2+\sigma_\mathrm{calib}^2}$, and propagated on to the intensity ratio using regular error propagation. Even though some lines were measured in a joint spectral setup, the full calibration uncertainty is included in the determination of all errors. In the cases where the ${}^{14}$N/${}^{15}$N ratios are derived from HCN isotopologues, the errors listed should  therefore be an upper limit on the uncertainty. 

\subsection{Optical depth considerations}\label{sec:optdepth}
\subsubsection{Hyperfine structure fitting for IRAS~16293A} 
The assumption that the lines, in particular H$^{13}$CN(3-2), are optically thin and hence their intensity is a direct measure of the number of molecules, needs to be verified. 

For IRAS~16293-2422, we have a H$^{13}$CN(1-0) spectrum \citep{Jorgensen04} from the SEST telescope\footnote{The Swedish-ESO 15m Submillimeter Telescope (SEST) was decommissioned in 2003.} with clearly resolved hyperfine structure (cf. Fig.~\ref{fig:iras16293_sest}, using a main beam efficiency of 0.75). The pointing is slightly different ($\alpha$ = 16:32:22.80, $\delta$ = $-$24:28:33.0) from our recent observations, but given the large beam the spectrum should still provide an estimate of the opacity. The three hyperfine transitions were fitted using the ``HFS method'' of the CLASS software to derive the opacity. The HFS routine fits Gaussians to all hyperfine components simultaneously under the assumption of a common excitation temperature, with relative positions and intensities provided by the user \citep[see the CLASS manual or appendix A of][for more details]{Hilyblant13a}. The best fit, illustrated in Fig.~\ref{fig:iras16293_sest}, yields an opacity for the main component of $\tau_\mathrm{main} = 0.274\pm0.405$ and hence the line is optically thin.

\placefigureSEST

\subsubsection{Radiative transfer modeling} 
The optical depth on higher excited lines can be larger than the opacity of H$^{13}$CN(1-0). Therefore, line radiative transfer models can be used as an alternative method to estimate the optical depth in the isotopologues and check the validity of Eq.~\ref{eqn:simpleratio}. The spherically symmetric source model from \citet{Schoier02} for IRAS~16293-2422, the R~CrA~IRS7B model with external irradiation from \citet{Lindberg12}, and the best fit model of OMC-3~MMS6 from \citet{Jorgensen06} are used as an input for the 1D version of the ``RATRAN'' radiative transfer code \citep{Hogerheijde00}. A comparison of the source models, i.e., density and temperature profiles, is shown in Fig.~\ref{fig:sourceprofiles}. We use molecular data for the HCN isotopologues from the LAMDA database\footnote{\url{http://home.strw.leidenuniv.nl/~moldata/}} \citep{Schoier05}. The transition frequencies, energy levels, and Einstein A coefficients in the LAMDA files are based on data from the CDMS catalog \citep{Mueller01,Mueller05}, and the collisional rate coefficients are scaled and extrapolated HCN-He collision rates from \citet{Green74}. Individual models were checked against newer collision rates from \citet{Dumouchel10} and the differences were found to be at the 5-10\% level in these cases, i.e., well within the calibration uncertainty.

The abundance profile is chosen according to the results by \citet{Jorgensen05apj}, i.e., a drop abundance profile accounting for freeze-out of the molecule at temperatures below a certain threshold and sufficient densities. Following \citet{Jorgensen05apj}, the freeze-out region for HCN in NGC~1333~IRAS~2A is defined as the part of the envelope with densities above $n > 7\times 10^{-4}~\mathrm{cm}^{-3}$ and temperatures below $T < 90~\mathrm{K}$, which corresponds to the region where CO and HCO$^{+}$ are also depleted. However, the density at which HCN is frozen out is not a fixed value but time and source dependent \citep{Jorgensen05freezeout}.\\

\placefigsourceprof

Therefore, models with three different freeze-out/evaporation temperatures ($30~\mathrm{K}$, $60~\mathrm{K}$, and $90~\mathrm{K}$) and various abundances were tested for IRAS~16293-2422. The density is higher than  $n > 7\times 10^{-4}~\mathrm{cm}^{-3}$ throughout the envelope in this model and a depletion zone was therefore not employed. A Doppler b-parameter of $db = 2.3~\mathrm{km}~\mathrm{s}^{-1}$ was assumed, corresponding to a line width (FWHM) of $\sim3.8~\mathrm{km}~\mathrm{s}^{-1}$. The model FWHM was chosen a bit lower than the observed line widths, because no additional velocity profile (e.g. infall) was used in the models.

The model with the abundance jump at $90~\mathrm{K}$ and $x_\mathrm{out} \approx 6\times10^{-12}$ (for $T < 90~\mathrm{K}$) and $x_\mathrm{in} \approx 1\times10^{-9}$ (where $T \geq 90~\mathrm{K}$) can reproduce the observed H$^{13}$CN(3-2 and 4-3) integrated intensities from IRAS~16293-2422 within a few percent. However, this model underproduces the HC$^{15}$N lines, in particular the $J = 4-3$ transition, by at least 20\% for H$^{13}$CN/HC$^{15}$N abundance ratios in the range of $2.6-6.3$, corresponding to a $^{14}$N/$^{15}$N ratio of $180-440$ assuming $^{12}\mathrm{C}/^{13}\mathrm{C} = 69$ \citep{Wilson99}. In addition, an abundance jump of more than two orders of magnitude is almost a factor of five higher than what was found in earlier work \citep{Jorgensen05apj,Brinch09}. Therefore, we also tested models where the jump occurs at $60~$K and $30~$K instead of $90~$K. For $60~$K, the H$^{13}$CN intensities are well modelled using $x_\mathrm{in} \approx 2\times10^{-10}$ and $x_\mathrm{out} \approx 8\times10^{-12}$. Using an HC$^{15}$N abundances of $x_\mathrm{in} \approx 7.7\times10^{-11}$ and $x_\mathrm{out} \approx 3.1\times10^{-12}$, corresponding to $^{14}$N/$^{15}$N = 180 as obtained using Eq.~\ref{eqn:simpleratio}, yields a very good fit to the observed integrated intensities of HC$^{15}$N. In contrast, using a solar ratio of $^{14}$N/$^{15}$N = 441 ($x_\mathrm{in} \approx 3.1\times10^{-11}$ and $x_\mathrm{out} \approx 1.3\times10^{-12}$) underproduces the line intensities by $\sim50\%$. For $30~$K, a good fit to the H$^{13}$CN lines is obtained using $x_\mathrm{in} \approx 4\times10^{-11}$ and $x_\mathrm{out} \approx 5\times10^{-12}$. Again, $^{14}\mathrm{N}/^{15}\mathrm{N} = 180$ ($x_\mathrm{in} \approx 1.5\times10^{-11}$ and $x_\mathrm{out} \approx 1.9\times10^{-12}$) gives an excellent fit to the HC$^{15}$N data, whereas a solar abundance ratio, $x_\mathrm{in} \approx 6.3\times10^{-12}$ and $x_\mathrm{out} \approx 7.8\times10^{-13}$, underestimates the observed integrated intensites by $\sim50\%$.

So while the $90~\mathrm{K}$ jump model, which becomes moderately optically thick for H$^{13}$CN ($\tau_{3-2} \approx 1.5$ towards line and source center), can reproduce the H$^{13}$CN lines, it cannot fit HC$^{15}$N for $^{14}$N/$^{15}$N ratios ranging from the most extreme values found from the simple analysis (Eq.~\ref{eqn:simpleratio}) to solar abundances. The optically thin models with the jump occuring at $60~$K and $30~$K on the other hand are also able to reproduce the HC$^{15}$N integrated intensities for $^{14}\mathrm{N}/^{15}\mathrm{N} = 180$, but not for solar abundances. Therefore, the radiative transfer results support the assumption that the line emission is optically thin, and that Eq.~\ref{eqn:simpleratio} is valid in this case.

For OMC-3~MMS6, a fairly well constrained model is also available. An abundance profile with $x \approx 8\times10^{-13}$ for $T < 90~\mathrm{K}$ and $n > 1\times 10^{6}~\mathrm{cm}^{-3}$ and $x \approx 7\times10^{-11}$ elsewhere yields integrated intensities that are consistent with the observed values  within the uncertainties ($\sim 20\%$). The corresponding opacities are in the range of $\tau \sim 0.5$ for $db = 0.6$ ($\mathrm{FWHM} \sim 1~\mathrm{km}~\mathrm{s}^{-1}$) and indicate that the H$^{13}$CN emission from this source is optically thin. For HC$^{15}$N(3-2), the integrated intensity can be reproduced within 20\% using models with $^{14}\mathrm{N}/^{15}\mathrm{N} = 370$ (as derived from Eq.~1) as well as $^{14}\mathrm{N}/^{15}\mathrm{N} = 441$ (solar ratio). As already concluded in Sec.~\ref{sec:ratios}, it is therefore not possible to distinguish between a slightly fractionated and a solar $^{14}\mathrm{N}/^{15}\mathrm{N}$ ratio for OMC-3~MMS6. The higher freeze-out density compared to IRAS~16293-2422 could indicate that this source is in a less evolved state. 

Because of the complex physical structure of R~CrA~IRS7B, including a disk \citep{Lindberg14b}, the spherically symmetric model is less reliable for this source than for IRAS~16293-2422 and OMC-3~MMS6. An agreement within 20\% of the modeled and the observed intensities can be obtained with an H$^{13}$CN abundance of $x \approx 2\times10^{-10}$ for $T < 90~\mathrm{K}$ and $n > 7\times 10^{4}~\mathrm{cm}^{-3}$ and $x \approx 2\times10^{-8}$ elsewhere. The optical depth in the central line channel and for a ray directly through the source center are $\tau \sim 1-3$. The observed HC$^{15}$N emission can be reproduced within the uncertainties with a range of $^{14}\mathrm{N}/^{15}\mathrm{N}$ ratios ($\sim 270-441$) because the model is not well constrained and only one HC$^{15}$N line is available. 
However, the required high abundances are an indication that the physical model underestimates the observed column density, which results in a higher optical depth. Compared to the IRAS~16293-2422 model, which has a similar outer radius, but higher densities, R~CrA~IRS7B is not expected to be significantly optically thick. Given the model uncertainties and the degeneracies in the parameter space, the simple analysis (Eq.~\ref{eqn:simpleratio}) is likely to provides the most reliable results for this source.

\section{Discussion}\label{sec:discussion}
At least two measurements of the ${}^{14}$N/${}^{15}$N ratio have been obtained in all three sources. For IRAS~16239A, the two measurements from the J=3-2 and 4-3 lines of H$^{13}$CN and HC$^{15}$N, ${}^{14}$N/${}^{15}$N $=163\pm20$ and ${}^{14}$N/${}^{15}$N $=190\pm38$, agree well. The ${}^{14}$N/${}^{15}$N $=242\pm32$ value from the HNC(3-2) isotopologues is different from the HCN values at the $1\sigma$ level. In the case of R~CrA~IRS7B, the HCN and HNC values are consistent within the error bars, including also the ratio derived from the data by \citet{Watanabe12}. The ${}^{14}$N/${}^{15}$N ratios from HCN and HNC in OMC-3~MMS6 agree within errors, but the uncertainties are larger in this source than in the two others. Overall, the agreement between different measurements of the same source is fairly good. The comparison to other measurements and literature values is discussed in the subsections below. 

Comparing Table~\ref{tab:ratios} with the temperatures in the outer regions of the spherically symmetric envelope models of the three sources in the sample, it is apparent that the lowest ${}^{14}$N/${}^{15}$N ratios are found in the source with the lowest temperature in the outermost regions, IRAS~16293A. It is followed by R~CrA~IRS7B, which has a somewhat elevated temperature because it is being irradiated by a nearby Herbig Be star, R~CrA \citep{Lindberg12}. The highest, almost solar, ${}^{14}$N/${}^{15}$N ratios are from OMC-3~MMS6, located in the active region of Orion and thus subject to a strong radiation field, which has the highest temperatures in the outer envelope. More quantitatively, this behavior is illustrated in Fig.~\ref{fig:temp_ratio}, showing the average of the ${}^{14}$N/${}^{15}$N ratio for each source versus the temperature at the projected radius of the beam. The APEX measurements listed in Table~\ref{tab:ratios} were averaged for each source and the errors bars reflect the spread of the measured ratios. The temperature was taken from the spherically symmetric source models, evaluated at the radius of the APEX beam ($\sim 24\arcsec$) at the corresponding source distance ($T = 16.5~$K for IRAS~16239A, $25.5~$K for R~CrA~IRS7B, and $37~$K for OMC-3~MMS6). A larger sample would be needed to draw any conclusions about whether or not there is a direct correlation between temperature and ${}^{15}$N-enhancement, but our result provides at least an indication for a temperature dependence of the ${}^{14}$N/${}^{15}$N ratio in these embedded protostars. 

\placefigtempratio

\subsection{Comparison with previous HCN and HNC measurements in the target sources}
Isotopologues of HCN and HNC in IRAS~16293-2422 were previously measured in the spectral surveys by \citet[][IRAM and JCMT]{Caux11} and \citet[][JCMT and CSO]{Blake94,vanDishoeck95}. The integrated intensities from the \citet{Caux11} and \citet{vanDishoeck95} surveys are listed in Table~\ref{tab:caux} in the appendix.
Note that the observations by \citet{Caux11} were pointed at the B component of the protobinary, whereas our observations are pointed directly at the A component. The two components are separated by $\sim~4\arcsec$ and thus lie both within the beam. However, in the 230~GHz band observations at the IRAM~30m, the flux of the A component is considerably attenuated \citep[see discussion in][]{Caux11}. The resulting ${}^{14}$N/${}^{15}$N ratios are, assuming a calibration uncertainty of 15\%, $276\pm74$ from the H$^{13}$CN(1-0) and HC$^{15}$N(1-0) intensities, $219\pm71$ for the 3-2 transitions, and $320\pm106$ for the 4-3 lines and thus comparable to our values. No H$^{15}$NC intensities are reported by \citet{Caux11}. 

From the integrated intensities of HC$^{15}$N(3-2) and H$^{13}$CN(3-2) measured by \citet{vanDishoeck95} we find ${}^{14}$N/${}^{15}$N = 131. From the fact that HC$^{15}$N is detected, they argue - presumably based on the assumption of a local interstellar medium value for the nitrogen isotope ratio \citep[][]{Wilson94} - that H$^{13}$CN is still slightly optically thick. However, from our model results and the fit to the hyperfine structure of H$^{13}$CN(1-0) we conclude that the lines are optically thin and that the assumption of a standard nitrogen isotope ratio likely does not hold because nitrogen is fractionated. 

R~CrA~IRS7B was surveyed in the 345~GHz spectral window by \citet{Watanabe12} using the ASTE telescope. The integrated intensities are $0.23\pm0.09~\mathrm{K}~\mathrm{km}~\mathrm{s}^{-1}$ for HC$^{15}$N(4-3), $0.95\pm0.10~\mathrm{K}~\mathrm{km}~\mathrm{s}^{-1}$ for H$^{13}$CN(4-3), $0.32\pm0.06~\mathrm{K}~\mathrm{km}~\mathrm{s}^{-1}$ for HN$^{13}$C(4-3), and $0.05\pm0.04~\mathrm{K}~\mathrm{km}~\mathrm{s}^{-1}$ for H$^{15}$NC(4-3). Based on exactly the same procedure as for our own measurements but with a calibration uncertainty of 19\% for ASTE \citep{Watanabe12}, the inferred ${}^{14}$N/${}^{15}$N ratios are $285\pm139$ from HCN and $442\pm382$ from HNC. The ratios derived from the work by \citet{Watanabe12} are consistent with our measurements, but the error bars are significantly larger than for our APEX observations. 

\citet{Schoier06} observed HCN(4-3) and HNC(4-3) with APEX in R~CrA~IRS7B, resulting in integrated intensities of $30.4~\mathrm{K}~\mathrm{km}~\mathrm{s}^{-1}$ and $11.3~\mathrm{K}~\mathrm{km}~\mathrm{s}^{-1}$, respectively. Their intensities are very similar to the values found from our observations of HCN(3-2) and HNC(3-2), which are $30.2~\mathrm{K}~\mathrm{km}~\mathrm{s}^{-1}$ and $14.2~\mathrm{K}~\mathrm{km}~\mathrm{s}^{-1}$. For optically thin HCN lines, similar line strengths for the 4--3 and 3--2 transitions are predicted by radiative transfer models if a constant abundance is assumed \citep[see also][for a discussion of the HCN abundance structure]{Jorgensen05apj}. Therefore, the similar intensities of the 4--3 and 3--2 transitions could be an indication for little variations in the HCN abundance with radius in this source, and thus no significant freeze-out of HCN.  

No HCN and HNC isotopologue measurements were found in the literature for OMC-3~MMS6. 

\subsection{Comparison with $^{14}$N/$^{15}$N ratios from the literature}
$^{14}$N/$^{15}$N ratios were previously measured in a selection of low-mass pre- and protostellar environments from a variety of molecular tracers. The literature results are summarized in Table~\ref{tab:prevnew}. The table clearly illustrates the discrepancy of $^{14}$N/$^{15}$N ratios derived from HCN and NH$_3$, which may be caused by different fractionation routes for nitriles and nitrogen hydrides \citep{Wirstroem12,Hilyblant13a,Hilyblant13b}. It also highlights the lack of a comprehensive set of $^{14}$N/$^{15}$N ratios from different tracers for a source, which makes the interpretation and comparison of individual measurements very challenging. For instance, \citet{Adande12} found a reasonable agreement between the $^{14}$N/$^{15}$N ratios inferred from HNC and CN. We find that the ratios derived from HCN and HNC are relatively similar. On the other hand, the prestellar core nitrogen isotope composition varies between HCN and CN \citep{Hilyblant13a,Hilyblant13b}, so it is not clear if there is any correlation between the $^{14}$N/$^{15}$N ratios from HCN, HNC, and CN despite the fact that they all belong to to the ``nitrile'' fractionation route \citep{Wirstroem12}. Moreover, the HNC values reported in \citet{Milam12} are substantially lower than any other measurement for reasons which are unclear. 

For IRAS~16293A, the inferred ratios from HCN (two transitions) and HNC (one measurement) fall into the range of the typical prestellar core HCN values \citep[][]{Hilyblant13a}.  
The similarity of the nitrogen isotope ratio in IRAS~16293A to the prestellar cores is an indication that either fractionation (not necessarily chemical) continues in the envelope of this YSO or that an enhanced $^{15}$N signature from the prestellar core phase has survived into the Class~0 phase. An enhanced radiation field seems to lower or even suppress nitrogen fractionation, as can be seen for the two sources with an enhanced external radiation field, R~CrA~IRS7B and OMC-3~MMS6. The $^{14}$N/$^{15}$N ratio in R~CrA~IRS7B lies at the upper end of the prestellar core values and the $^{14}$N/$^{15}$N ratio from OMC-3~MMS6 is even higher, outside the range spanned by the prestellar cores.

\subsection{Comparison to the local interstellar medium}
To decide whether a source shows an enhancement of $^{15}$N in a molecule, one would like to compare the $^{14}$N/$^{15}$N ratio from the source to a reference value of the interstellar medium (ISM). Different $^{14}$N/$^{15}$N measurements of the local interstellar medium are available from the literature.  
\citet{Wilson99} list $^{14}$N/$^{15}$N$ = 388\pm32$ for the local ISM \citep[updated from the earlier value of $^{14}$N/$^{15}$N$ = 450\pm22$,][]{Wilson94} based on the HCN surveys data from \citet[][for the northern sources]{Wannier81} and \citet[][for the southern sky]{Dahmen95} as well as the NH$_3$ survey by \citet{Guesten85}. However, \citet{Wielen97} argue that the ISM value should be $^{14}\mathrm{N}/^{15}\mathrm{N} = 414\pm32$, taking into account that the Sun formed at a smaller galactocentric radius that its current location.  
A local ISM value of $^{14}$N/$^{15}$N$ = 290\pm40$ is reported by \citet{Adande12} from measurements of HNC and CN in warm clouds and the $^{12}$C/$^{13}$C ratios from \citet{Milam05}. While the local value from the work by \citet{Adande12} is inferred from a linear fit to the data throughout the Galaxy, the local ratio reported in \citet{Wilson99} is the average of 8 sources close to the solar galactic distance. The \citet{Wilson99} and \citet{Adande12} results are not consistent within error bars, and the difference presumably arises from the different carbon isotope ratios ($^{12}$C/$^{13}$C) assumed for the double isotope method in the case of HCN and HNC. The $^{12}$C/$^{13}$C ratio from CN and CO isotopologues, used in \citet{Adande12}, are generally lower than those obtained from H$_2$CO isotopologues, included in \citet{Wilson99}. However, the local $^{12}$C/$^{13}$C ratios from \citet[$69\pm6$ from CO and H$_2$CO,][]{Wilson99} and \citet[][$68\pm15$ from CO, CN, and H$_2$CO]{Milam05} agree well, but the values for individual sources close to the solar neighborhood from \citet{Milam05} are partly lower, which is then transferred onto the $^{14}$N/$^{15}$N ratios from HNC \citep{Adande12}. The \citet{Adande12} result agrees with the average value of $^{14}$N/$^{15}$N$ = 237_{-21}^{+27}$ measured in two diffuse clouds along the line of sight toward a strong extragalactic source from absorption spectroscopy of HCN \citep{Lucas98}. 
It was however noticed that there is a significant scatter of the data points from the Galactic isotope surveys, and thus it is unclear how meaningful the comparison of the obtained $^{14}$N/$^{15}$N ratios to an average local ISM value is to decide whether or not the nitrogen isotopes are fractionated in a specific source. Moreover, the local ISM value is also subject to change, as galactic evolution models by \citet{Romano03} for instance predict that the $^{14}$N/$^{15}$N ratio of the ISM increases with galactocentric distance and decreases with time. 

Our IRAS~16293A results from HCN lie outside the uncertainty range of both the \citet{Adande12} and \citet{Wilson99} local ISM values, while the ratio inferred from HNC lies within the uncertainties of \citet{Adande12}, but clearly outside \citet{Wilson99}, as do the R~CrA HCN and HNC ratios. The HCN measurement from OMC-3~MMS6 is agreement with both ISM values, while the ratio from HNC, which is the most uncertain of all data points, is consistent with the \citet{Wilson99} ISM result but falls above the one from \citet{Adande12}. Hence, from the comparison with ISM measurements, the nitrogen isotopes in IRAS~16293A show indications for fractionation, and fractionation may possibly also occur in R~CrA~IRS7B, while OMC-3~MMS6 shows no evidence for fractionation if the comparison with a local ISM value is meaningful. The ${}^{12}\mathrm{C}/{}^{13}\mathrm{C}$ ratio should also be measured for our sources to check and account for any deviations from the standard.

\subsection{Comparison to solar system values}
A direct comparison of the nitrogen isotope composition with that of solar system bodies is difficult as possible mechanisms for a separation of the protostellar material into a $^{15}$N-poor gas phase and $^{15}$N-rich solids are not yet understood. The freeze-out of $^{15}$N-substituted molecular species, in particular ammonia, onto dust grains and their incorporation into the ice mantles is believed to be important \citep{Rodgers08a,Wirstroem12}, but the detailed processes are still unclear. Existing observations of the nitrogen isotope composition in star-forming regions do not point at a large nitrogen isotope fractionation in ammonia \citep{Gerin09,Daniel13}. \citet{Gerin09} and \citet{Bizzocchi13} even argue for a depletion of $^{15}$N in NH$_3$ and N$_2$H$^{+}$, respectively, in the prestellar core L1544. 
The comparison is further complicated by the fact that the chemical carrier of the elevated $^{15}$N-signature or the functional group carrying the isotopic anomaly in primitive solar system matter has not yet been identified. Furthermore, the cosmomaterials that can nowadays be sampled may represent only a minor reservoir of the presolar nebula. It may have also underdone significant processing during the evolution of the solar system, such as heating or aqueous alteration, which could affect the isotopic signatures \citep[e.g.,][]{Alexander07}. 

Bulk ${}^{15}$N-enrichments in meteorites are comparable to the values inferred from star-forming regions. However, meteoritic ${}^{15}$N-hotspots feature enrichments exceeding the one most extreme values for prestellar cores \citep{Hilyblant13a} or our protostar sample. Within errors, the cometary ratios are also in agreement with our HCN results for IRAS~16293A, but lower than the OMC-3~MMS6 and R~CrA~IRS7B results. Overall, the $^{14}$N/$^{15}$N ratios in primitive solar system matter, i.e., meteorites, comets, and IDPs, are comparable to the ones measured around prestellar cores and protostars (cf. comparison in Fig.~\ref{fig:nitrogen_ratios}), but the ambiguities on both the meteoritic and the astrochemical side do not yet allow concluding whether or not the ${}^{15}$N-signature in meteorites was directly inherited from a ${}^{15}$N-enriched precursor molecule in the presolar nebula or whether it is the result of a secondary process. 

\subsection{Chemical fractionation or photochemistry?}

A possible explanation for the isotopic variations among our sample could be the different amounts of external irradiation the sources are exposed to. IRAS~16293A, being a relatively isolated source, has the lowest temperatures in the outer envelope, whereas OMC-3~MMS6 and R~CrA~IRS7B are both exposed to an enhanced radiation field, resulting in warmer outer envelope temperatures \citep{Jorgensen06,Lindberg12}. There is a tentative trend toward a decreasing ${}^{15}$N-enhancement with increasing outer envelope temperature in our sample, but the statistics is too small to draw any conclusions. If the trend can be confirmed, it would argue in favor of the chemical fractionation scenario over isotope-selective photochemistry. If photodissociation of N$_2$ was the process dominating the nitrogen isotope composition, the highest ${}^{15}$N-enrichments should be present in the two externally irradiated sources, not IRAS~16293A. However, the single-dish observations only provide a spatially averaged value of the ${}^{14}$N/${}^{15}$N ratio. Because fractionation may not occur everywhere in the envelope, and because the mechanism that dominates the nitrogen isotope composition may even vary throughout the envelope depending on physical parameters such as temperature, irradiation, and opacity, spatially resolved observations are clearly needed to study potential spatial variations in ${}^{14}$N/${}^{15}$N ratio. For instance, photochemistry could determine the ${}^{14}$N/${}^{15}$N ratio in the outermost and maybe innermost parts close to the protostar, where the radiation field is intense, whereas chemical fractionation is expected to be important in cold, shielded regions. Our understanding of the evolution of nitrogen isotopic composition during star- and planet formation would also benefit from a more comprehensive data set with nitrogen isotope ratios measured from a census of molecular tracers and sources in different evolutionary stages. Nonetheless, our results indicate that enhanced ${}^{15}$N-values in HCN and HNC are present not just in the prestellar stage, but also around at least some Class~0 protostars. Whether fractionation is still ongoing in these sources or just preserved from the precedent prestellar core phase and what mechanism is causing the ${}^{15}$N-enrichment in HCN and HNC is not yet clear and needs to be further investigated.

\section{Conclusions}\label{sec:conclusions}
We have measured the ${}^{14}$N/${}^{15}$N isotope ratio in the three Class~0 protostars IRAS~16293A, R~CrA~IRS7B, and OMC-3~MMS6 from ${}^{15}$N- and ${}^{13}$C-substituted isotopologues of HCN and HNC, observed with the APEX telescope. Optically thin emission, the same excitation temperature for both isotopologues, and a standard carbon isotope ratio of ${}^{12}\mathrm{C}/{}^{13}\mathrm{C} = 69$ \citep{Wilson99,Milam05} were assumed to derive the ${}^{14}$N/${}^{15}$N ratios directly from the integrated line intensities. The assumption of optically thin emission is supported by the results from H${}^{13}$CN(1-0) hyperfine structure fitting for IRAS~16293A and by non-LTE line radiative transfer modeling for all three sources. 

The ratios derived from the HCN and HNC isotopologues are relatively similar, but there is significant variation between the tree sources. The values for IRAS~16293A (${}^{14}\mathrm{N}/{}^{15}\mathrm{N} \sim 160-240$) are in the range of typical prestellar core values \citep[${}^{14}\mathrm{N}/{}^{15}\mathrm{N} \sim 140-360$,][]{Hilyblant13a}, whereas the results for R~CrA~IRS7B (${}^{14}\mathrm{N}/{}^{15}\mathrm{N} \sim 260-290$) indicate that the nitrogen isotopes in this source are less fractionated. For OMC-3~MMS6 (${}^{14}\mathrm{N}/{}^{15}\mathrm{N} \sim 370-460$), a solar and hence unfractionated nitrogen isotope composition cannot be excluded. The differences between the sources possibly stem from the different outer envelope temperatures, presumably caused by external irradiation heating the envelope. 

Our results indicate that fractionation of nitrogen isotopes in HCN and HNC either continues in the Class~0 stage or is at least preserved from the prestellar core phase. Additional tracers should also be studied to learn more about what molecules carry enhanced ${}^{15}\mathrm{N}$ and could be responsible for the ${}^{15}\mathrm{N}$-enrichment observed in solar system materials.

Chemical fractionation seems a more likely scenario than N$_2$ self-shielding to explain the nitrogen isotope measurements in our data set because of a tentative trend toward a decreasing ${}^{14}\mathrm{N}/{}^{15}\mathrm{N}$ with decreasing envelope temperature. However, spatially resolved observations are crucial to pinpoint which mechanism causes the observed ${}^{15}$N-enrichment in some Class~0 protostars.

\bibliographystyle{aa}
\bibliography{mybib}

\begin{thebibliography}{98}
\expandafter\ifx\csname natexlab\endcsname\relax\def\natexlab#1{#1}\fi

\bibitem[{{Abbas} {et~al.}(2004){Abbas}, {LeClair}, {Owen}, {Conrath},
  {Flasar}, {Kunde}, {Nixon}, {Achterberg}, {Bjoraker}, {Jennings}, {Orton}, \&
  {Romani}}]{Abbas04}
{Abbas}, M.~M., {LeClair}, A., {Owen}, T., {et~al.} 2004, \apj, 602, 1063

\bibitem[{{Adande} \& {Ziurys}(2012)}]{Adande12}
{Adande}, G.~R. \& {Ziurys}, L.~M. 2012, \apj, 744, 194

\bibitem[{{Al{\'e}on}(2010)}]{Aleon10}
{Al{\'e}on}, J. 2010, \apj, 722, 1342

\bibitem[{{Alexander} {et~al.}(2007){Alexander}, {Fogel}, {Yabuta}, \&
  {Cody}}]{Alexander07}
{Alexander}, C.~M.~{\relax O\'{}D}., {Fogel}, M., {Yabuta}, H., \& {Cody},
  G.~D. 2007, \gca, 71, 4380

\bibitem[{{Anders} \& {Grevesse}(1989)}]{Anders89}
{Anders}, E. \& {Grevesse}, N. 1989, \gca, 53, 197

\bibitem[{{Bertin} {et~al.}(2013){Bertin}, {Fayolle}, {Romanzin}, {Poderoso},
  {Michaut}, {Philippe}, {Jeseck}, {{\"O}berg}, {Linnartz}, \&
  {Fillion}}]{Bertin13}
{Bertin}, M., {Fayolle}, E.~C., {Romanzin}, C., {et~al.} 2013, \apj, 779, 120

\bibitem[{{Birck}(2004)}]{Birck04}
{Birck}, J.~L. 2004, \rmg, 55, 25

\bibitem[{{Bizzocchi} {et~al.}(2010){Bizzocchi}, {Caselli}, \&
  {Dore}}]{Bizzocchi10}
{Bizzocchi}, L., {Caselli}, P., \& {Dore}, L. 2010, \aap, 510, L5+

\bibitem[{{Bizzocchi} {et~al.}(2013){Bizzocchi}, {Caselli}, {Leonardo}, \&
  {Dore}}]{Bizzocchi13}
{Bizzocchi}, L., {Caselli}, P., {Leonardo}, E., \& {Dore}, L. 2013, \aap, 555,
  A109

\bibitem[{{Blake} {et~al.}(1994){Blake}, {van Dishoeck}, {Jansen}, {Groesbeck},
  \& {Mundy}}]{Blake94}
{Blake}, G.~A., {van Dishoeck}, E.~F., {Jansen}, D.~J., {Groesbeck}, T.~D., \&
  {Mundy}, L.~G. 1994, \apj, 428, 680

\bibitem[{{Bockel{\'e}e-Morvan} {et~al.}(2008){Bockel{\'e}e-Morvan}, {Biver},
  {Jehin}, {Cochran}, {Wiesemeyer}, {Manfroid}, {Hutsem{\'e}kers}, {Arpigny},
  {Boissier}, {Cochran}, {Colom}, {Crovisier}, {Milutinovic}, {Moreno},
  {Prochaska}, {Ramirez}, {Schulz}, \& {Zucconi}}]{Bockelee08}
{Bockel{\'e}e-Morvan}, D., {Biver}, N., {Jehin}, E., {et~al.} 2008, \apjl, 679,
  L49

\bibitem[{{B{\"o}hlke} {et~al.}(2005){B{\"o}hlke}, {de Laeter}, {De
  Bi\`{e}vre}, {Hidaka}, {Peiser}, {Rosman}, \& {Taylor}}]{Bohlke05}
{B{\"o}hlke}, J.~K., {de Laeter}, J.~R., {De Bi\`{e}vre}, P., {et~al.} 2005, J.
  Phys. Chem. Ref. Data, 34, 57

\bibitem[{{Bonal} {et~al.}(2010){Bonal}, {Huss}, {Krot}, {Nagashima}, {Ishii},
  \& {Bradley}}]{Bonal10}
{Bonal}, L., {Huss}, G.~R., {Krot}, A.~N., {et~al.} 2010, \gca, 74, 6590

\bibitem[{{Boogert} {et~al.}(2002){Boogert}, {Hogerheijde}, {Ceccarelli},
  {Tielens}, {van Dishoeck}, {Blake}, {Latter}, \& {Motte}}]{Boogert02}
{Boogert}, A.~C.~A., {Hogerheijde}, M.~R., {Ceccarelli}, C., {et~al.} 2002,
  \apj, 570, 708

\bibitem[{{Briani} {et~al.}(2009){Briani}, {Gounelle}, {Marrocchi},
  {Mostefaoui}, {Leroux}, {Quirico}, \& {Meibom}}]{Briani09}
{Briani}, G., {Gounelle}, M., {Marrocchi}, Y., {et~al.} 2009, Proceedings of
  the National Academy of Sciences, 106

\bibitem[{{Brinch} {et~al.}(2009){Brinch}, {J{\o}rgensen}, \&
  {Hogerheijde}}]{Brinch09}
{Brinch}, C., {J{\o}rgensen}, J.~K., \& {Hogerheijde}, M.~R. 2009, \aap, 502,
  199

\bibitem[{{Busemann} {et~al.}(2006){Busemann}, {Young}, {O'D.~Alexander},
  {Hoppe}, {Mukhopadhyay}, \& {Nittler}}]{Busemann06}
{Busemann}, H., {Young}, A.~F., {O'D.~Alexander}, C.~M., {et~al.} 2006,
  Science, 312, 727

\bibitem[{{Caux} {et~al.}(2011){Caux}, {Kahane}, {Castets}, {Coutens},
  {Ceccarelli}, {Bacmann}, {Bisschop}, {Bottinelli}, {Comito}, {Helmich},
  {Lefloch}, {Parise}, {Schilke}, {Tielens}, {van Dishoeck}, {Vastel},
  {Wakelam}, \& {Walters}}]{Caux11}
{Caux}, E., {Kahane}, C., {Castets}, A., {et~al.} 2011, \aap, 532, A23

\bibitem[{{Clayton}(2002)}]{Clayton02}
{Clayton}, R.~N. 2002, \nat, 415, 860

\bibitem[{{Connelly} {et~al.}(2012){Connelly}, {Bizzarro}, {Krot}, {Nordlund},
  {Wielandt}, \& {Ivanova}}]{Connelly12}
{Connelly}, J.~N., {Bizzarro}, M., {Krot}, A.~N., {et~al.} 2012, Science, 338,
  651

\bibitem[{{Coplen} {et~al.}(1992){Coplen}, {Krouse}, \&
  {B{\"o}hlke}}]{Coplen92}
{Coplen}, T.~B., {Krouse}, H.~R., \& {B{\"o}hlke}, J.~K. 1992, Pure Appl. Chem,
  64, 907

\bibitem[{{Dahmen} {et~al.}(1995){Dahmen}, {Wilson}, \& {Matteucci}}]{Dahmen95}
{Dahmen}, G., {Wilson}, T.~L., \& {Matteucci}, F. 1995, \aap, 295, 194

\bibitem[{{Daniel} {et~al.}(2013){Daniel}, {G{\'e}rin}, {Roueff}, {Cernicharo},
  {Marcelino}, {Lique}, {Lis}, {Teyssier}, {Biver}, \&
  {Bockel{\'e}e-Morvan}}]{Daniel13}
{Daniel}, F., {G{\'e}rin}, M., {Roueff}, E., {et~al.} 2013, \aap, 560, A3

\bibitem[{{Dumke} \& {Mac-Auliffe}(2010)}]{Dumke10}
{Dumke}, M. \& {Mac-Auliffe}, F. 2010, in Society of Photo-Optical
  Instrumentation Engineers (SPIE) Conference Series, Vol. 7737

\bibitem[{{Dumouchel} {et~al.}(2010){Dumouchel}, {Faure}, \&
  {Lique}}]{Dumouchel10}
{Dumouchel}, F., {Faure}, A., \& {Lique}, F. 2010, \mnras, 406, 2488

\bibitem[{{Fletcher} {et~al.}(2014){Fletcher}, {Greathouse}, {Orton}, {Irwin},
  {Mousis}, {Sinclair}, \& {Giles}}]{Fletcher14}
{Fletcher}, L.~N., {Greathouse}, T.~K., {Orton}, G.~S., {et~al.} 2014, \icarus,
  238, 170

\bibitem[{{Floss} {et~al.}(2006){Floss}, {Stadermann}, {Bradley}, {Dai},
  {Bajt}, {Graham}, \& {Lea}}]{Floss06}
{Floss}, C., {Stadermann}, F.~J., {Bradley}, J.~P., {et~al.} 2006, \gca, 70,
  2371

\bibitem[{{Fouchet} {et~al.}(2004){Fouchet}, {Irwin}, {Parrish}, {Calcutt},
  {Taylor}, {Nixon}, \& {Owen}}]{Fouchet04}
{Fouchet}, T., {Irwin}, P.~G.~J., {Parrish}, P., {et~al.} 2004, \icarus, 172,
  50

\bibitem[{{Fouchet} {et~al.}(2000){Fouchet}, {Lellouch}, {B{\'e}zard},
  {Encrenaz}, {Drossart}, {Feuchtgruber}, \& {de Graauw}}]{Fouchet00}
{Fouchet}, T., {Lellouch}, E., {B{\'e}zard}, B., {et~al.} 2000, \icarus, 143,
  223

\bibitem[{{Gerin} {et~al.}(2009){Gerin}, {Marcelino}, {Biver}, {Roueff},
  {Coudert}, {Elkeurti}, {Lis}, \& {Bockel{\'e}e-Morvan}}]{Gerin09}
{Gerin}, M., {Marcelino}, N., {Biver}, N., {et~al.} 2009, \aap, 498, L9

\bibitem[{{Green} \& {Thaddeus}(1974)}]{Green74}
{Green}, S. \& {Thaddeus}, P. 1974, \apj, 191, 653

\bibitem[{{Guesten} \& {Ungerechts}(1985)}]{Guesten85}
{Guesten}, R. \& {Ungerechts}, H. 1985, \aap, 145, 241

\bibitem[{{G{\"u}sten} {et~al.}(2006){G{\"u}sten}, {Nyman}, {Schilke},
  {Menten}, {Cesarsky}, \& {Booth}}]{Guesten06}
{G{\"u}sten}, R., {Nyman}, L.~{\AA}., {Schilke}, P., {et~al.} 2006, \aap, 454,
  L13

\bibitem[{{Hashizume} {et~al.}(2000){Hashizume}, {Chaussidon}, {Marty}, \&
  {Robert}}]{Hashizume00}
{Hashizume}, K., {Chaussidon}, M., {Marty}, B., \& {Robert}, F. 2000, Science,
  290, 1142

\bibitem[{{Heays} {et~al.}(2014){Heays}, {Visser}, {Gredel}, {Ubachs}, {Lewis},
  {Gibson}, \& {van Dishoeck}}]{Heays14}
{Heays}, A.~N., {Visser}, R., {Gredel}, R., {et~al.} 2014, \aap, 562, A61

\bibitem[{{Hily-Blant} {et~al.}(2013{\natexlab{a}}){Hily-Blant}, {Bonal},
  {Faure}, \& {Quirico}}]{Hilyblant13a}
{Hily-Blant}, P., {Bonal}, L., {Faure}, A., \& {Quirico}, E.
  2013{\natexlab{a}}, \icarus, 223, 582

\bibitem[{{Hily-Blant} {et~al.}(2013{\natexlab{b}}){Hily-Blant}, {Pineau des
  For{\^e}ts}, {Faure}, {Le Gal}, \& {Padovani}}]{Hilyblant13b}
{Hily-Blant}, P., {Pineau des For{\^e}ts}, G., {Faure}, A., {Le Gal}, R., \&
  {Padovani}, M. 2013{\natexlab{b}}, \aap, 557, A65

\bibitem[{{Hoffman} {et~al.}(1979){Hoffman}, {Hodges}, {McElroy}, {Donahue}, \&
  {Kolpin}}]{Hoffman79}
{Hoffman}, J.~H., {Hodges}, R.~R., {McElroy}, M.~B., {Donahue}, T.~M., \&
  {Kolpin}, M. 1979, Science, 205, 49

\bibitem[{{Hogerheijde} \& {van der Tak}(2000)}]{Hogerheijde00}
{Hogerheijde}, M.~R. \& {van der Tak}, F.~F.~S. 2000, \aap, 362, 697

\bibitem[{{Hutsem{\'e}kers} {et~al.}(2005){Hutsem{\'e}kers}, {Manfroid},
  {Jehin}, {Arpigny}, {Cochran}, {Schulz}, {St{\"u}we}, \&
  {Zucconi}}]{Hutsemekers05}
{Hutsem{\'e}kers}, D., {Manfroid}, J., {Jehin}, E., {et~al.} 2005, \aap, 440,
  L21

\bibitem[{{Ikeda} {et~al.}(2002){Ikeda}, {Hirota}, \& {Yamamoto}}]{Ikeda02}
{Ikeda}, M., {Hirota}, T., \& {Yamamoto}, S. 2002, \apj, 575, 250

\bibitem[{{Jehin} {et~al.}(2009){Jehin}, {Manfroid}, {Hutsem{\'e}kers},
  {Arpigny}, \& {Zucconi}}]{Jehin09}
{Jehin}, E., {Manfroid}, J., {Hutsem{\'e}kers}, D., {Arpigny}, C., \&
  {Zucconi}, J.-M. 2009, Earth Moon and Planets, 105, 167

\bibitem[{{Johnstone} {et~al.}(2003){Johnstone}, {Boonman}, \& {van
  Dishoeck}}]{Johnstone03}
{Johnstone}, D., {Boonman}, A.~M.~S., \& {van Dishoeck}, E.~F. 2003, \aap, 412,
  157

\bibitem[{{J{\o}rgensen} {et~al.}(2005{\natexlab{a}}){J{\o}rgensen}, {Bourke},
  {Myers}, {Sch{\"o}ier}, {van Dishoeck}, \& {Wilner}}]{Jorgensen05apj}
{J{\o}rgensen}, J.~K., {Bourke}, T.~L., {Myers}, P.~C., {et~al.}
  2005{\natexlab{a}}, \apj, 632, 973

\bibitem[{{J{\o}rgensen} {et~al.}(2011){J{\o}rgensen}, {Bourke}, {Nguyen
  Luong}, \& {Takakuwa}}]{Jorgensen11}
{J{\o}rgensen}, J.~K., {Bourke}, T.~L., {Nguyen Luong}, Q., \& {Takakuwa}, S.
  2011, \aap, 534, A100

\bibitem[{{J{\o}rgensen} {et~al.}(2008){J{\o}rgensen}, {Johnstone}, {Kirk},
  {Myers}, {Allen}, \& {Shirley}}]{Jorgensen08}
{J{\o}rgensen}, J.~K., {Johnstone}, D., {Kirk}, H., {et~al.} 2008, \apj, 683,
  822

\bibitem[{{J{\o}rgensen} {et~al.}(2006){J{\o}rgensen}, {Johnstone}, {van
  Dishoeck}, \& {Doty}}]{Jorgensen06}
{J{\o}rgensen}, J.~K., {Johnstone}, D., {van Dishoeck}, E.~F., \& {Doty}, S.~D.
  2006, \aap, 449, 609

\bibitem[{{J{\o}rgensen} {et~al.}(2004){J{\o}rgensen}, {Sch{\"o}ier}, \& {van
  Dishoeck}}]{Jorgensen04}
{J{\o}rgensen}, J.~K., {Sch{\"o}ier}, F.~L., \& {van Dishoeck}, E.~F. 2004,
  \aap, 416, 603

\bibitem[{{J{\o}rgensen} {et~al.}(2005{\natexlab{b}}){J{\o}rgensen},
  {Sch{\"o}ier}, \& {van Dishoeck}}]{Jorgensen05freezeout}
{J{\o}rgensen}, J.~K., {Sch{\"o}ier}, F.~L., \& {van Dishoeck}, E.~F.
  2005{\natexlab{b}}, \aap, 435, 177

\bibitem[{{Junk} \& {Svec}(1958)}]{Junk58}
{Junk}, G. \& {Svec}, H.~J. 1958, \gca, 14, 234

\bibitem[{{Kerridge}(1993)}]{Kerridge93}
{Kerridge}, J.~F. 1993, Reviews of Geophysics, 31, 423

\bibitem[{{Li} {et~al.}(2013){Li}, {Heays}, {Visser}, {Ubachs}, {Lewis},
  {Gibson}, \& {van Dishoeck}}]{Li13}
{Li}, X., {Heays}, A.~N., {Visser}, R., {et~al.} 2013, \aap, 555, A14

\bibitem[{{Liang} {et~al.}(2007){Liang}, {Heays}, {Lewis}, {Gibson}, \&
  {Yung}}]{Liang07b}
{Liang}, M.-C., {Heays}, A.~N., {Lewis}, B.~R., {Gibson}, S.~T., \& {Yung},
  Y.~L. 2007, \apjl, 664, L115

\bibitem[{{Lindberg} \& {J{\o}rgensen}(2012)}]{Lindberg12}
{Lindberg}, J.~E. \& {J{\o}rgensen}, J.~K. 2012, \aap, 548, A24

\bibitem[{{Lindberg} {et~al.}(2014){Lindberg}, {J{\o}rgensen}, {Brinch},
  {Haugb{\o}lle}, {Bergin}, {Harsono}, {Persson}, {Visser}, \&
  {Yamamoto}}]{Lindberg14b}
{Lindberg}, J.~E., {J{\o}rgensen}, J.~K., {Brinch}, C., {et~al.} 2014, \aap,
  566, A74

\bibitem[{{Linke} {et~al.}(1977){Linke}, {Goldsmith}, {Wannier}, {Wilson}, \&
  {Penzias}}]{Linke77}
{Linke}, R.~A., {Goldsmith}, P.~F., {Wannier}, P.~G., {Wilson}, R.~W., \&
  {Penzias}, A.~A. 1977, \apj, 214, 50

\bibitem[{{Lis} {et~al.}(2010){Lis}, {Wootten}, {Gerin}, \& {Roueff}}]{Lis10}
{Lis}, D.~C., {Wootten}, A., {Gerin}, M., \& {Roueff}, E. 2010, \apjl, 710, L49

\bibitem[{{Loinard} {et~al.}(2000){Loinard}, {Castets}, {Ceccarelli},
  {Tielens}, {Faure}, {Caux}, \& {Duvert}}]{Loinard00}
{Loinard}, L., {Castets}, A., {Ceccarelli}, C., {et~al.} 2000, \aap, 359, 1169

\bibitem[{{Lucas} \& {Liszt}(1998)}]{Lucas98}
{Lucas}, R. \& {Liszt}, H. 1998, \aap, 337, 246

\bibitem[{{Lyons} {et~al.}(2009){Lyons}, {Bergin}, {Ciesla}, {Davis}, {Desch},
  {Hashizume}, \& {Lee}}]{Lyons09}
{Lyons}, J.~R., {Bergin}, E.~A., {Ciesla}, F.~J., {et~al.} 2009, \gca, 73, 4998

\bibitem[{{Manfroid} {et~al.}(2009){Manfroid}, {Jehin}, {Hutsem{\'e}kers},
  {Cochran}, {Zucconi}, {Arpigny}, {Schulz}, {St{\"u}we}, \&
  {Ilyin}}]{Manfroid09}
{Manfroid}, J., {Jehin}, E., {Hutsem{\'e}kers}, D., {et~al.} 2009, \aap, 503,
  613

\bibitem[{{Mariotti}(1983)}]{Mariotti83}
{Mariotti}, A. 1983, \nat, 303, 685

\bibitem[{{Marty} {et~al.}(2011){Marty}, {Chaussidon}, {Wiens}, {Jurewicz}, \&
  {Burnett}}]{Marty11}
{Marty}, B., {Chaussidon}, M., {Wiens}, R.~C., {Jurewicz}, A.~J.~G., \&
  {Burnett}, D.~S. 2011, Science, 332, 1533

\bibitem[{{Mathew} \& {Marti}(2001)}]{Mathew01}
{Mathew}, K.~J. \& {Marti}, K. 2001, \jgr, 106, 1401

\bibitem[{{McKeegan} {et~al.}(2006){McKeegan}, {Al{\'e}on}, {Bradley},
  {Brownlee}, {Busemann}, {Butterworth}, {Chaussidon}, {Fallon}, {Floss},
  {Gilmour}, {Gounelle}, {Graham}, {Guan}, {Heck}, {Hoppe}, {Hutcheon}, {Huth},
  {Ishii}, {Ito}, {Jacobsen}, {Kearsley}, {Leshin}, {Liu}, {Lyon}, {Marhas},
  {Marty}, {Matrajt}, {Meibom}, {Messenger}, {Mostefaoui}, {Mukhopadhyay},
  {Nakamura-Messenger}, {Nittler}, {Palma}, {Pepin}, {Papanastassiou},
  {Robert}, {Schlutter}, {Snead}, {Stadermann}, {Stroud}, {Tsou}, {Westphal},
  {Young}, {Ziegler}, {Zimmermann}, \& {Zinner}}]{McKeegan06}
{McKeegan}, K.~D., {Al{\'e}on}, J., {Bradley}, J., {et~al.} 2006, Science, 314,
  1724

\bibitem[{{Meibom} {et~al.}(2007){Meibom}, {Krot}, {Robert}, {Mostefaoui},
  {Russell}, {Petaev}, \& {Gounelle}}]{Meibom07}
{Meibom}, A., {Krot}, A.~N., {Robert}, F., {et~al.} 2007, \apjl, 656, L33

\bibitem[{{Messenger}(2000)}]{Messenger00}
{Messenger}, S. 2000, \nat, 404, 968

\bibitem[{{Milam} \& {Charnley}(2012)}]{Milam12}
{Milam}, S.~N. \& {Charnley}, S.~B. 2012, in Lunar and Planetary Inst.
  Technical Report, Vol.~43, Lunar and Planetary Institute Science Conference
  Abstracts, 2618

\bibitem[{{Milam} {et~al.}(2005){Milam}, {Savage}, {Brewster}, {Ziurys}, \&
  {Wyckoff}}]{Milam05}
{Milam}, S.~N., {Savage}, C., {Brewster}, M.~A., {Ziurys}, L.~M., \& {Wyckoff},
  S. 2005, \apj, 634, 1126

\bibitem[{{M{\"u}ller} {et~al.}(2005){M{\"u}ller}, {Schl{\"o}der}, {Stutzki},
  \& {Winnewisser}}]{Mueller05}
{M{\"u}ller}, H.~S.~P., {Schl{\"o}der}, F., {Stutzki}, J., \& {Winnewisser}, G.
  2005, Journal of Molecular Structure, 742, 215

\bibitem[{{M{\"u}ller} {et~al.}(2001){M{\"u}ller}, {Thorwirth}, {Roth}, \&
  {Winnewisser}}]{Mueller01}
{M{\"u}ller}, H.~S.~P., {Thorwirth}, S., {Roth}, D.~A., \& {Winnewisser}, G.
  2001, \aap, 370, L49

\bibitem[{{Mumma} \& {Charnley}(2011)}]{Mumma11}
{Mumma}, M.~J. \& {Charnley}, S.~B. 2011, \araa, 49, 471

\bibitem[{{Niemann} {et~al.}(2005){Niemann}, {Atreya}, {Bauer}, {Carignan},
  {Demick}, {Frost}, {Gautier}, {Haberman}, {Harpold}, {Hunten}, {Israel},
  {Lunine}, {Kasprzak}, {Owen}, {Paulkovich}, {Raulin}, {Raaen}, \&
  {Way}}]{Niemann05}
{Niemann}, H.~B., {Atreya}, S.~K., {Bauer}, S.~J., {et~al.} 2005, \nat, 438,
  779

\bibitem[{{Owen} {et~al.}(2001){Owen}, {Mahaffy}, {Niemann}, {Atreya}, \&
  {Wong}}]{Owen01}
{Owen}, T., {Mahaffy}, P.~R., {Niemann}, H.~B., {Atreya}, S., \& {Wong}, M.
  2001, \apjl, 553, L77

\bibitem[{{Persson} {et~al.}(2013){Persson}, {J{\o}rgensen}, \& {van
  Dishoeck}}]{Persson13}
{Persson}, M.~V., {J{\o}rgensen}, J.~K., \& {van Dishoeck}, E.~F. 2013, \aap,
  549, L3

\bibitem[{{Pickett} {et~al.}(1998){Pickett}, {Poynter}, {Cohen}, {Delitsky},
  {Pearson}, \& {M{\"u}ller}}]{Pickett98}
{Pickett}, H.~M., {Poynter}, R.~L., {Cohen}, E.~A., {et~al.} 1998, \jqsrt, 60,
  883

\bibitem[{{Pineda} {et~al.}(2012){Pineda}, {Maury}, {Fuller}, {Testi},
  {Garc{\'{\i}}a-Appadoo}, {Peck}, {Villard}, {Corder}, {van Kempen}, {Turner},
  {Tachihara}, \& {Dent}}]{Pineda12}
{Pineda}, J.~E., {Maury}, A.~J., {Fuller}, G.~A., {et~al.} 2012, \aap, 544, L7

\bibitem[{{Rodgers} \& {Charnley}(2008{\natexlab{a}})}]{Rodgers08b}
{Rodgers}, S.~D. \& {Charnley}, S.~B. 2008{\natexlab{a}}, \apj, 689, 1448

\bibitem[{{Rodgers} \& {Charnley}(2008{\natexlab{b}})}]{Rodgers08a}
{Rodgers}, S.~D. \& {Charnley}, S.~B. 2008{\natexlab{b}}, \mnras, 385, L48

\bibitem[{{Romano} \& {Matteucci}(2003)}]{Romano03}
{Romano}, D. \& {Matteucci}, F. 2003, \mnras, 342, 185

\bibitem[{{Rousselot} {et~al.}(2014){Rousselot}, {Pirali}, {Jehin}, {Vervloet},
  {Hutsem{\'e}kers}, {Manfroid}, {Cordier}, {Martin-Drumel}, {Gruet},
  {Arpigny}, {Decock}, \& {Mousis}}]{Rousselot14}
{Rousselot}, P., {Pirali}, O., {Jehin}, E., {et~al.} 2014, \apjl, 780, L17

\bibitem[{{Sch{\"o}ier} {et~al.}(2006){Sch{\"o}ier}, {J{\o}rgensen},
  {Pontoppidan}, \& {Lundgren}}]{Schoier06}
{Sch{\"o}ier}, F.~L., {J{\o}rgensen}, J.~K., {Pontoppidan}, K.~M., \&
  {Lundgren}, A.~A. 2006, \aap, 454, L67

\bibitem[{{Sch{\"o}ier} {et~al.}(2002){Sch{\"o}ier}, {J{\o}rgensen}, {van
  Dishoeck}, \& {Blake}}]{Schoier02}
{Sch{\"o}ier}, F.~L., {J{\o}rgensen}, J.~K., {van Dishoeck}, E.~F., \& {Blake},
  G.~A. 2002, \aap, 390, 1001

\bibitem[{{Sch{\"o}ier} {et~al.}(2005){Sch{\"o}ier}, {van der Tak}, {van
  Dishoeck}, \& {Black}}]{Schoier05}
{Sch{\"o}ier}, F.~L., {van der Tak}, F.~F.~S., {van Dishoeck}, E.~F., \&
  {Black}, J.~H. 2005, \aap, 432, 369

\bibitem[{{Shinnaka} {et~al.}(2014){Shinnaka}, {Kawakita}, {Kobayashi},
  {Nagashima}, \& {Boice}}]{Shinnaka14}
{Shinnaka}, Y., {Kawakita}, H., {Kobayashi}, H., {Nagashima}, M., \& {Boice},
  D.~C. 2014, \apjl, 782, L16

\bibitem[{{Takakuwa} {et~al.}(2007){Takakuwa}, {Ohashi}, {Bourke}, {Hirano},
  {Ho}, {J{\o}rgensen}, {Kuan}, {Wilner}, \& {Yeh}}]{Takakuwa07}
{Takakuwa}, S., {Ohashi}, N., {Bourke}, T.~L., {et~al.} 2007, \apj, 662, 431

\bibitem[{{Tennekes} {et~al.}(2006){Tennekes}, {Harju}, {Juvela}, \&
  {T{\'o}th}}]{Tennekes06}
{Tennekes}, P.~P., {Harju}, J., {Juvela}, M., \& {T{\'o}th}, L.~V. 2006, \aap,
  456, 1037

\bibitem[{{Terzieva} \& {Herbst}(2000)}]{Terzieva00}
{Terzieva}, R. \& {Herbst}, E. 2000, \mnras, 317, 563

\bibitem[{{van Dishoeck} {et~al.}(1995){van Dishoeck}, {Blake}, {Jansen}, \&
  {Groesbeck}}]{vanDishoeck95}
{van Dishoeck}, E.~F., {Blake}, G.~A., {Jansen}, D.~J., \& {Groesbeck}, T.~D.
  1995, \apj, 447, 760

\bibitem[{{Vassilev} {et~al.}(2008){Vassilev}, {Meledin}, {Lapkin}, {Belitsky},
  {Nystr{\"o}m}, {Henke}, {Pavolotsky}, {Monje}, {Risacher}, {Olberg},
  {Strandberg}, {Sundin}, {Fredrixon}, {Ferm}, {Desmaris}, {Dochev},
  {Pantaleev}, {Bergman}, \& {Olofsson}}]{Vassilev08}
{Vassilev}, V., {Meledin}, D., {Lapkin}, I., {et~al.} 2008, \aap, 490, 1157

\bibitem[{{Wannier} {et~al.}(1981){Wannier}, {Linke}, \& {Penzias}}]{Wannier81}
{Wannier}, P.~G., {Linke}, R.~A., \& {Penzias}, A.~A. 1981, \apj, 247, 522

\bibitem[{{Watanabe} {et~al.}(2012){Watanabe}, {Sakai}, {Lindberg},
  {J{\o}rgensen}, {Bisschop}, \& {Yamamoto}}]{Watanabe12}
{Watanabe}, Y., {Sakai}, N., {Lindberg}, J.~E., {et~al.} 2012, \apj, 745, 126

\bibitem[{{Wielen} \& {Wilson}(1997)}]{Wielen97}
{Wielen}, R. \& {Wilson}, T.~L. 1997, \aap, 326, 139

\bibitem[{{Wilson}(1999)}]{Wilson99}
{Wilson}, T.~L. 1999, Reports on Progress in Physics, 62, 143

\bibitem[{{Wilson} \& {Rood}(1994)}]{Wilson94}
{Wilson}, T.~L. \& {Rood}, R. 1994, \araa, 32, 191

\bibitem[{{Wirstr{\"o}m} {et~al.}(2012){Wirstr{\"o}m}, {Charnley}, {Cordiner},
  \& {Milam}}]{Wirstroem12}
{Wirstr{\"o}m}, E.~S., {Charnley}, S.~B., {Cordiner}, M.~A., \& {Milam}, S.~N.
  2012, \apjl, 757, L11

\bibitem[{{Womack} {et~al.}(1992){Womack}, {Ziurys}, \& {Wyckoff}}]{Womack92}
{Womack}, M., {Ziurys}, L.~M., \& {Wyckoff}, S. 1992, \apj, 387, 417

\bibitem[{{Wong} {et~al.}(2013){Wong}, {Atreya}, {Mahaffy}, {Franz},
  {Malespin}, {Trainer}, {Stern}, {Conrad}, {Manning}, {Pepin}, {Becker},
  {McKay}, {Owen}, {Navarro-Gonz{\'a}lez}, {Jones}, {Jakosky}, \&
  {Steele}}]{Wong13}
{Wong}, M.~H., {Atreya}, S.~K., {Mahaffy}, P.~N., {et~al.} 2013, \grl, 40, 6033

\end{thebibliography}

\begin{acknowledgements}
The authors thank the referee for constructive comments that helped to improve the article. Research for this project was supported by a Danish National Research Foundation Centre of Excellence grant to M.B. (grant number DNRF97) and from a Lundbeck Foundation Junior Group Leader Fellowship to J.K.J. M.B. further acknowledges funding from the European Research Council under ERC Consolidator grant agreement 616027-STARDUST2ASTEROIDS. The research leading to these results has received funding from the European Commission Seventh Framework Programme (FP/2007-2013) under grant agreement No 283393 (RadioNet3). The authors would like to thank the Onsala/APEX staff for carrying out the observations. Per Bergman and the Onsala/APEX staff are also thanked for help assessing potential contamination from the off position. 
\end{acknowledgements}

\Online

\begin{appendix}
\section{Observational details} \label{sec:obsdetails}

\begin{figure}[h!]
 \centering
 \resizebox{0.75\hsize}{!}{\includegraphics{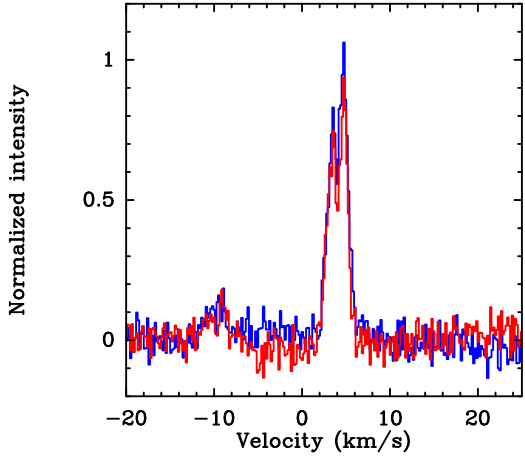}}
 \caption{Comparison of the H${}^{13}$CO$^+$(3-2) line from IRAS~16293A observed with APEX. The spectra resulting from phase 1 (corresponding to off position on one side of the source in symmetric wobbler switching mode) and phase 2 (off position on the other side) are shown in blue and red, respectively. The spectra have been baseline subtracted and normalized to the average peak intensity of the two spectra. Data from 2012-04-01.}
 \label{fig:iras16293_h13coplus_offcomparison}
\end{figure}

\begin{figure}[h!]
 \centering
 \resizebox{0.75\hsize}{!}{\includegraphics{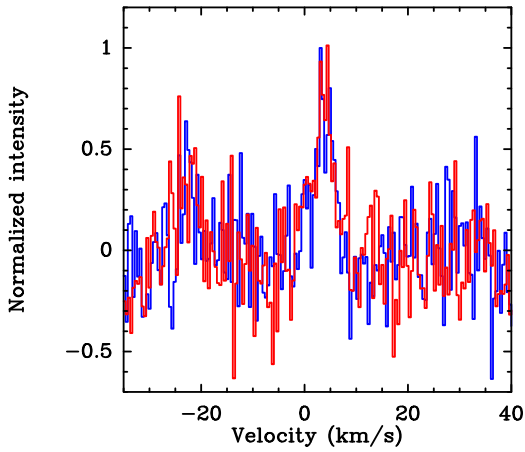}}
 \caption{Like Fig~\ref{fig:iras16293_h13coplus_offcomparison} but for H${}^{13}$CN(3-2). Includes a sinusoidal baseline for standing wave removal.}
 \label{fig:iras16293_h13cn_offcomparison}
\end{figure}

\begin{figure}[h!]
 \centering
 \resizebox{0.75\hsize}{!}{\includegraphics{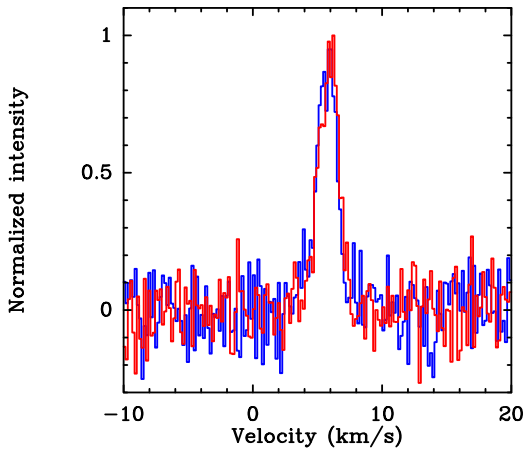}}
 \caption{Like Fig~\ref{fig:iras16293_h13coplus_offcomparison} but for R~CrA~IRS7B. Data from 2012-09-27.}
 \label{fig:rcrairs7b_h13coplus_offcomparison}
\end{figure}

\begin{figure}
 \centering
 \resizebox{0.75\hsize}{!}{\includegraphics{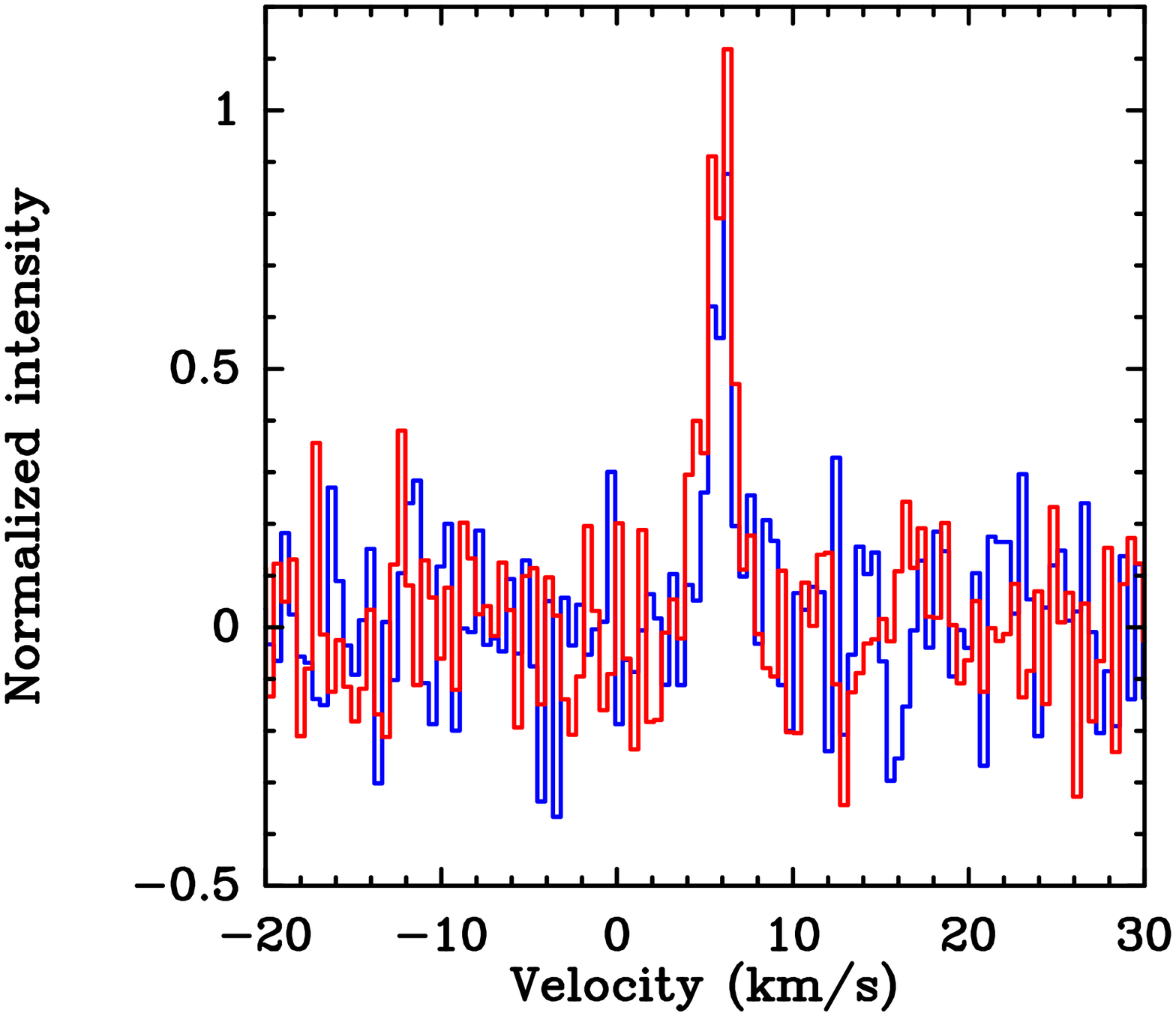}}
 \caption{Like Fig~\ref{fig:rcrairs7b_h13coplus_offcomparison} but for H${}^{13}$CN(3-2).}
 \label{fig:rcrairs7b_h13cn_offcomparison}
\end{figure}

\begin{figure}
 \centering
 \resizebox{0.75\hsize}{!}{\includegraphics{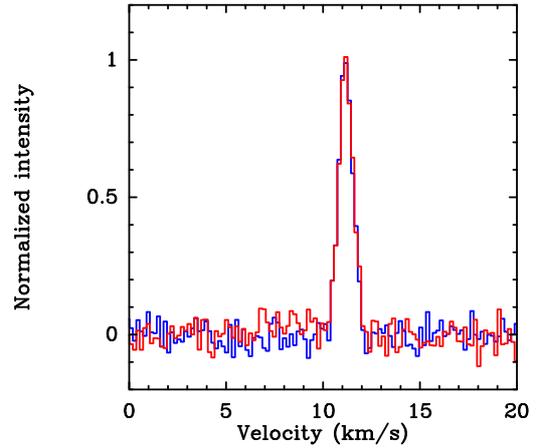}}
 \caption{Like Fig~\ref{fig:iras16293_h13coplus_offcomparison} but for OMC-3~MMS6. Data from 2012-11-22.}
 \label{fig:omc3mms6_h13coplus_offcomparison}
\end{figure}

\begin{figure}
 \centering
 \resizebox{0.75\hsize}{!}{\includegraphics{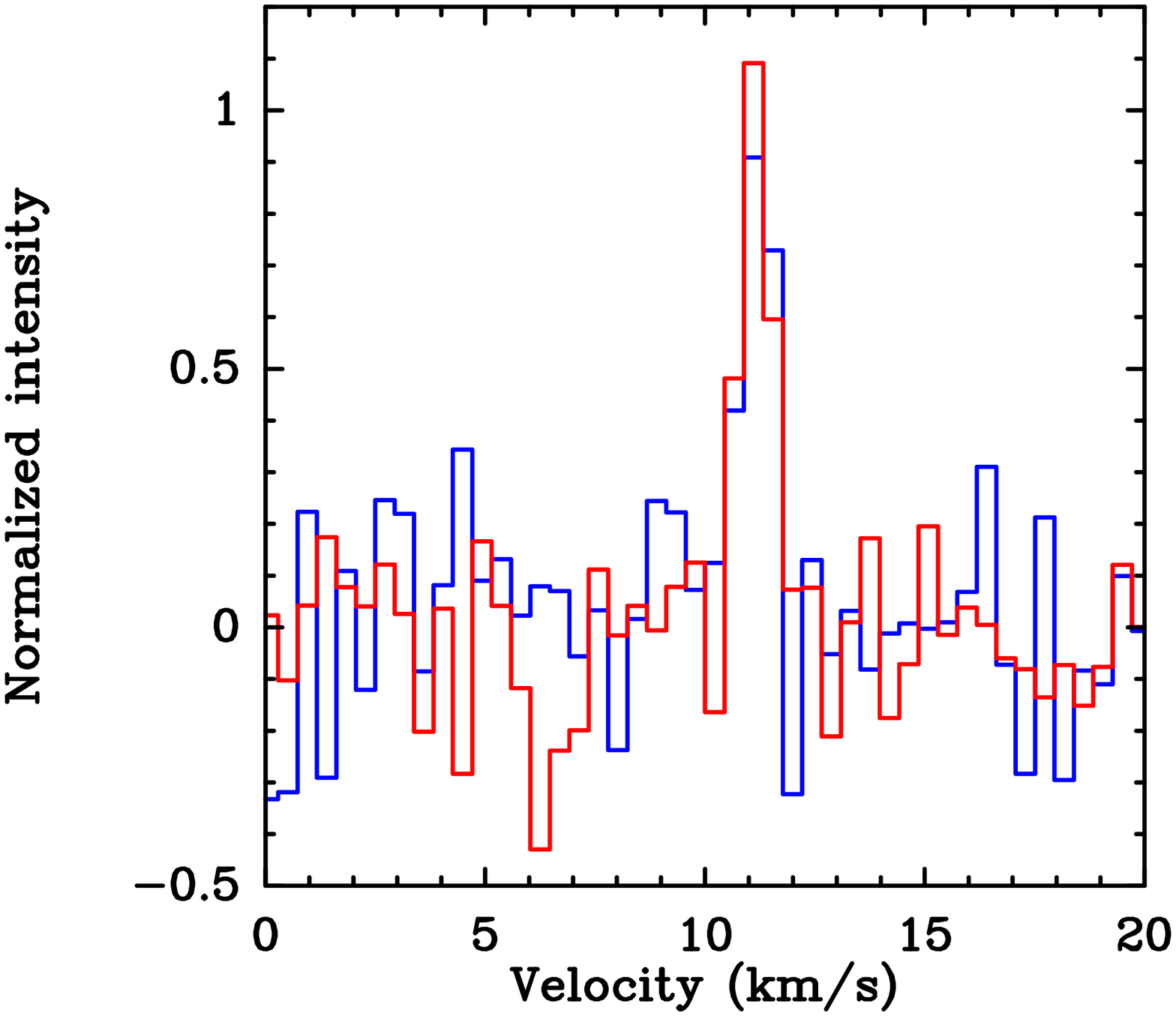}}
 \caption{Like Fig~\ref{fig:omc3mms6_h13coplus_offcomparison} but for H${}^{13}$CN(3-2).}
 \label{fig:omc3mms6_h13cn_offcomparison}
\end{figure}

\placetableobsdetail

\section{Literature values}

\placetableSolarSyst

\placetablecaux

\end{appendix}

\end{document}